\documentstyle[12pt,aaspp4]{article}

\begin{document}

\title{Arecibo \ion{H}{1} Mapping of a Large Sample of Dwarf Irregular Galaxies}

\author{G. Lyle Hoffman}
\affil{Dept. of Physics, Lafayette College}

\author{E.E. Salpeter}
\affil{Center for Radiophysics and Space Research, Cornell University}

\author{B. Farhat}
\affil{Dept. of Physics, Massachusetts Institute of Technology}

\author{T. Roos}
\affil{Dept. of Physics, Cornell University}

\author{H. Williams}
\affil{Dept. of Physics and Astronomy, University of Delaware}

\and

\author{G. Helou}
\affil{Infrared Processing and Analysis Center, Caltech}

\begin{abstract}

Neutral hydrogen mapping of 70 dwarf irregular (Sm, Im and BCD) galaxies is reported, with position-velocity contour maps (and a few contour maps in R.A. and Dec.)  presented for those resolved by the Arecibo beam.
The galaxies were selected either from the {\it Virgo Cluster Catalog}, from similarly identified field galaxies, or from a distance-limited sample within the Arecibo declination range.
We do not find any isolated dwarfs with a larger \ion{H}{1} to optical radius ratio than DDO 154; the ``protogalaxy'' HI 1225+01 (Giovanelli \& Haynes \markcite{GH89} 1989) continues to be a unique object among dwarfs that have been mapped in \ion{H}{1}.
For all dwarfs with significant rotation we are able to determine the sense of the spin.
For a number of better-resolved dwarfs we are able to determine rotation curves, in most cases extending well beyond the last measured point in available synthesis array maps.
Correlations among the several measures of galaxy size and mass are studied; in  Paper II (Salpeter \& Hoffman \markcite{SH96} 1996) we combine these data with those for the set of all available mapped dwarf irregular galaxies and for mapped spirals spanning a similar range of redshifts to investigate variations in Tully-Fisher relations and in surface densities as functions of galaxy size and luminosity or mass.


\end{abstract}

\section{Introduction}

Dwarf irregular galaxies are a very heterogeneous lot, ranging from ImV systems of such low surface brightness that they can only just be discerned on deep photographic plates to BCD or \ion{H}{2} galaxies that are evidently in the throes of intense star formation (Sandage \& Binggeli \markcite{SB84} 1984).
Spiral galaxies also have a wide spread in surface brightness (Sandage, Binggeli \& Tammann \markcite{SBT85} 1985; Roberts \& Haynes \markcite{RH94} 1994; Sprayberry et al. \markcite{SBIB95} 1995; McGaugh, Schombert \& Bothun \markcite{MSB95} 1995), and one question of interest is to what extent the dwarf irregulars are simply lower-mass analogues of the spiral systems and to what extent the irregulars form a distinct population (Hoffman, Helou \& Salpeter \markcite{HHS88} 1988; Zwaan et al. \markcite{ZvdHdBM95} 1995).
While it is conventional to define ``dwarf'' galaxies to be fainter than some chosen absolute magnitude, for our purposes it is convenient to distinguish on the basis of morphology alone so that throughout this paper "dwarf irregulars" will refer to galaxies of type Sdm, Sm, Im and BCD as defined by Sandage \& Binggeli \markcite{SB84} (1984), regardless of size or luminosity, while "spirals" include types Sb through Sd.

The neutral hydrogen reservoirs in dwarf irregulars (and spirals) are important for studies of star formation (Arimoto \& Tarrab \markcite{AT90} 1990; van der Hulst et al. \markcite{vdHSSBMdB93} 1993; Gavazzi \markcite{G93} 1993; Taylor et al. \markcite{TBPS94} 1994) and the age of these systems (Hodge \markcite{H89} 1989 and references therein; Tyson \& Scalo \markcite{TS88} 1988; Schombert et al. \markcite{SBIM90} 1990; van den Bergh \markcite{vdB94} 1994).
The recent discoveries of \ion{H}{1} 1225+01, an \ion{H}{1} cloud with a very sparse stellar component (Giovanelli \& Haynes \markcite{GH89} 1989; Salzer et al. \markcite{SSMGH91} 1991), and star-poor \ion{H}{1} companions of actively star-forming \ion{H}{2} galaxies (Taylor, Brinks \& Skillman \markcite{TBS93} 1993, \markcite{TBS95} 1995) has rekindled debate over the possibility that there are large numbers of dwarf-sized gas clouds still waiting to form their first stars (Fisher \& Tully \markcite{FT81} 1981; Hoffman et al. \markcite{HHSL89} 1989a; Briggs \markcite{B90} 1990; Weinberg et al. \markcite{WSGvG91} 1991; Hoffman, Lu \& Salpeter \markcite{HLS92} 1992; Briggs \& Rao \markcite{BR93} 1993; Hoffman \markcite{H94} 1994).
Overall dwarfs outnumber larger galaxies by a wide margin, making them important tracers of the large-scale mass distribution and --- if we can learn to determine reliable inclinations and disentangle the contributions of ordered rotation and random motions to the profile widths --- Tully-Fisher distance markers (Hoffman, Helou \& Salpeter \markcite{HHS88} 1988; Schneider et al. \markcite{STMW90} 1990; Gavazzi \markcite{G93} 1993; Sprayberry et al. \markcite{SBIB95} 1995; Zwaan et al. \markcite{ZvdHdBM95} 1995).

In most cases, especially for dwarf irregulars, the \ion{H}{1} gas reservoir extends to larger radii than the visible stars so that the rotation curve of the galaxy can be traced to larger radii in \ion{H}{1} than in optical spectra.
This makes the \ion{H}{1} an important probe of dark matter halos around galaxies (e.g., Puche \& Carignan \markcite{PC91} 1991; Begeman, Broeils \& Sanders \markcite{BBS91} 1991; Ashman \markcite{A92} 1992).
As an alternative to dark matter as an explanation for the dynamics of galaxies and galaxy clusters, Milgrom (\markcite{Mi83a}1983a,\markcite{Mi83b}b,\markcite{Mi83c}c; see also Sanders \markcite{S90} 1990 and references therein) proposed altering Newtonian dynamics in the extremely weak field limit.
Milgrom's version of modified Newtonian dynamics, MOND, lends itself well to tests against galaxy rotation curves, and the most stringent tests are provided by the lowest accelerations encountered --- the outermost parts of dwarf galaxies (Milgrom \markcite{Mi94} 1994; Lo, Sargent \& Young \markcite{LSY93} 1993 and others).

We began our program of mapping dwarf galaxies at Arecibo
\footnote{The Arecibo Observatory is part of the National Astronomy and Ionosphere Center, which is operated by Cornell University under a management agreement with the National Science Foundation.}
with the intention of identifying objects with extended rotation curves to explore constraints on dark matter candidates and MOND.
\ion{H}{1} 1225+01 and the \ion{H}{1} companions of \ion{H}{2} galaxies, discovered after most of our mapping was complete, provide {\em ex post facto} motivation:  our mapping provides constraints on the frequency with which such systems occur.
While synthesis telescopes (the Very Large Array and Westerbork) offer finer spatial resolution of the inner parts of the \ion{H}{1} disks, their deficit of short baselines deprives them of sensitivity to low-density gas in the outskirts unless mosaicing techniques are employed; and they cannot match Arecibo's velocity resolution.
Consequently Arecibo offers a more economical way to identify extended halos, especially for galaxies with \ion{H}{1} signals that are not particularly strong, and complements synthesis mappings with more detail of the outer envelope. 
In all, we have mapped 70 dwarfs to date; our most significant conclusion is that \ion{H}{1} disks that exceed the Holmberg diameter by more than a factor of six are quite rare.
DDO 154 (previously reported by Krumm \& Burstein \markcite{KB84} 1984; Carignan \& Freeman \markcite{CF88} 1988; Carignan, Beaulieu \& Freeman \markcite{CBF90} 1990) is the only isolated dwarf in our sample which is so extended; we discussed it and two dwarfs distended by interaction with companions in Hoffman et al. \markcite{HLSFLR93} (1993).
A second isolated example, \ion{H}{1} 1225+01, was discovered by Giovanelli \& Haynes \markcite{GH89} (1989) after our observing was complete.
However, we now have a large database with which to tackle the several other questions that dwarf galaxies can address.
We will for the most part defer discussion of dark matter constraints and MOND to forthcoming papers; here we present the data, discuss individual cases along with the statistics of extended \ion{H}{1} envelopes and extremely star-deficient gas clouds, and explore correlations among the various indicators of galaxy size:  \ion{H}{1} profile width, optical and \ion{H}{1} diameters, and luminosity.
In Paper II (Salpeter \& Hoffman \markcite{SH96} 1996) we  incorporate data from the literature to extend those studies to a larger sample of dwarfs and to investigate the relationship between dwarfs and spirals.

In Sect. 2 we discuss how the dwarfs we have mapped were selected.  
Our general observing strategies are explained in Sect. 3, and various corrections made to the data are detailed along with the results in Sect. 4.  
Approximately one quarter of the galaxies are resolved by the Arecibo beam; we discuss these in detail in Sect. 5.  
Comparisons are made to synthesis array mapping of several of our galaxies in Sect. 6, where we also discuss a number of special cases.  
We end with a general discussion in Sect. 7 and  a summary in Sect. 8. 

\section{Sample Selection}

Our total sample was selected in three lots, with some inevitable
overlap of criteria.  We shall call these lots ``Leo/Virgo sample,''
``Field survey sample,'' and ``Nearby dwarfs sample;'' and for most purposes
we will merge the three samples or select subsamples without regard to
lot boundaries.

The Leo/Virgo sample grew out of our extensive \ion{H}{1} survey of dwarf
irregular galaxies in the Virgo Cluster area (Hoffman et al. \markcite{HHSGS87} 1987
--- hereafter HHSGS, and \markcite{HWSSB89} 1989c --- hereafter HWSSB).  
From the catalog of 293 dwarf irregular galaxies in Binggeli,
Sandage \& Tammann \markcite{BST85} (1985 --- hereafter BST) and 16 from Ferguson \& 
Sandage \markcite{FS90} (1990) we drew a total of 45 to map.  
For the most part these
were the galaxies with the strongest central beam emission; however, we
augmented the sample with a number of fainter sources to test whether
diffuse \ion{H}{1} envelopes were more extended relative to the optical
extent than were the more dense \ion{H}{1} reservoirs.  
A few galaxies were
added simply because they fell at the edge of the Virgo window and could
be mapped before other sources entered the beam, or after they left.
The resulting sample represents essentially all dwarf irregular
morphologies found in BST, weighted somewhat in favor of the more
intense sources.

The Field survey sample was similarly selected from the Binggeli,
Tarenghi \& Sandage \markcite{BTS90} (1990 --- hereafter BTS) field dwarf survey.  
Our \ion{H}{1} detection
survey of the late-type dwarfs in this catalog is reported in HWSSB.  
Again
we selected mainly the more intense sources for mapping, with some
attention to the morphological mix; and a few sources were mapped
because they fell in a right ascension window that was sparsely
populated with other survey candidates.  In all, we mapped 14 of the 42
BTS dwarfs reported in HWSSB.

The Nearby dwarfs sample was intended to produce a complete sample of
mapped dwarf galaxies within the Arecibo declination range, out to a
specified distance from the Milky Way.  In the allotted telescope time,
we were able to extend that radius to 6 Mpc, for a total of 17 galaxies
(four of which were mapped as part of the Field survey and one as part
of the Virgo survey).  The Field survey adds another galaxy within
7 Mpc, but there are another two or three dwarfs between 6 and 7 Mpc
that have not yet been mapped at Arecibo.  

In sum, we present Arecibo mapping for 70 dwarf irregular galaxies among
these three samples.  Of these, 20 non-interacting galaxies are
sufficiently resolved for us to deduce something about their rotation
curves; we will refer to these 20 as the ``Resolved dwarf sample.''

\section{Observing Strategies}

For the most part we used the Arecibo dual circular 21cm feed in these observations, since we anticipated weak signals and small extents so that sidelobe removal (see Section 4) would not be difficult.  
The one exception was DDO 154, which is so extended that sidelobe contamination of the central profiles by gas $6\arcmin\;$ further out is significant, greatly compounding the sidelobe decontamination process.  
For this galaxy, after some preliminary work with the circular feed, we turned to the flat feed which has much less obtrusive sidelobes (Hoffman et al. \markcite{HLSFLR93} 1993).  
Our mapping of IC 1613 would have benefitted from use of the flat feed for similar reasons, although we have reasonable confidence in our iterative sidelobe removal in this case.

Since the profiles were known (from central beam observations) to be narrow, we were able to use frequency-switching with the galaxy's signal in both frequency buffers for all dual circular feed mapping.  
This allowed for a reduction in observing time by a factor of 2 from position-switched observations.  In most cases we mapped with 2 km $\rm s^{-1} \;$ channel spacing with either no smoothing (if the signal-to-noise ratio was already $> 10$) or quintic spline smoothing (which does not bias profile widths).  
Therefore our velocity resolution is $\sim$2 km $\rm s^{-1} \;$ in the final spectra.

For most galaxies we mapped in a simple cross pattern with $1.\arcmin9$ beam spacing.  
If a major axis position angle was known (most often from UGC [Nilson \markcite{UGC} 1973]), we aligned one arm of the cross with the major axis; otherwise we used a simple cross aligned with the ordinal directions.  
For galaxies that proved to be quite extended, or if we had kinematical evidence that the position angle did not adequately represent the \ion{H}{1} disk, we added beams along axes bisecting the original ones.  
The two galaxies which we knew in advance to be extended were mapped in a hexagonal lattice pattern, with 3.\arcmin9 spacing for DDO 154 and 5.\arcmin7 spacing for IC 1613.

We did not, in general, seek a thorough mapping of each dwarf in the Leo/Virgo sample.  We were more interested in obtaining a cursory mapping for a large number of objects, in search of any with large dark matter halos.  
Consequently the signal-to-noise ratio in many of the maps for that sample is less than ideal.  
Our maps in the Nearby and Field samples have more consistently high signal-to-noise ratios at all points except the outermost, and even those have sufficiently long integrations that the rms noise is $\sim$1 mJy (after spline smoothing) in spite of the narrow velocity resolution.  
For DDO 154 in particular we sought to extend the rotation curve as far as possible, so that we obtained rms noise $\sim$0.6 mJy at the outermost points (Hoffman et al. \markcite{HLSFLR93} 1993).

\section{Data Corrections and Interpretation}

\subsection{Sidelobe removal}

The Arecibo dual circular 21cm feed has a pronounced $-11$dB first sidelobe which
peaks at a radius of $\sim 6\arcmin \;$ and has an integrated response equal to
$\sim$80\% of the main beam.  For this reason it is often shunned for
mapping work in favor of the less sensitive flat feed.  
For galaxies of very large extent this remains the only recourse at
present; but for galaxies that are only a few beamwidths in extent it is
generally possible to effectively remove the sidelobe contributions from
each spectrum and thus take advantage of the superior sensitivity of the
dual circular feed.  In addition, since these dwarf galaxies have quite
narrow velocity widths, it is possible to speed up the mapping by
another factor $\sim$2
by frequency-switching with the galaxy's signal in both buffers (which
is inadvisable for the flat feed due to the complicated ripple in its gain
vs. frequency curve).

Sidelobe removal is accomplished iteratively using the Arecibo spectrum analysis
package ANALYZ.  We assume, as in Krumm \& Salpeter \markcite{KS79} (1979), Helou
et al. \markcite{HGSK81} (1981), and Schneider et al. \markcite{SHST86} (1986),
that the sidelobe ring can be divided into octants,
each with an integrated response of about 10\% of the main beam.  For the
first pass correction to spectrum A, 10\% of the spectrum B recorded at
the position of the sidelobe is subtracted (channel by channel) from
spectrum A.  For a sufficiently edge-on galaxy, only points along the
major axis need be considered; for more nearly face-on galaxies it is
essential to have data along both major and minor axes (and in some
cases at additional off-axis points).  Then the emission falling in each
octant of the ring can be interpolated between observed points.  If the
points $6\arcmin\;$ from the galaxy's center have sufficiently weak flux, so
that 10\% of that flux is negligible compared to the flux observed at the
center, a single pass is usually sufficient; for more extended galaxies
a second pass is made:  10\% of the first-pass spectrum B is subtracted
from the original spectrum A.  If necessary, a third pass is made; but
in such cases the sidelobe contributions to the central flux are so
large and uncertain that it is best to use the flat feed instead (as we
have done for one galaxy in our sample, DDO 154).  In cases where
sidelobes are drawn from the opposite end of the major axis, or from
points further in along a rotation curve that is rising (or falling)
appreciably, we have added confidance in the sidelobe removal procedure
since the sidelobe contributions fall at a different velocity than the
flux in the mainbeam.  In practice this was true for all but one or two
of the galaxies we mapped.

\subsection{Neutral hydrogen diameters and masses}

Galaxy-wide \ion{H}{1} masses and effective diameters $D_{H,e}$ at $1/e$ times the central flux (for sufficiently resolved galaxies) were calculated as
described in Helou, Hoffman \& Salpeter \markcite{HHS84} (1984 --- hereafter HHS84) and 
Hoffman et al. \markcite{HLHSW89} (1989b).
In short, we fit the observed fluxes at the several beam positions to an \ion{H}{1} disk modeled by a flat central distribution with exponential edges; for simplicity the radius of the flat center and the scale length of the exponential edges are taken to be the same for each galaxy.
The diameter $D_{H,e}$ is taken to be the diameter of the best fitting model at which the \ion{H}{1} emission has fallen to $1/e$ times the emission at the center.
The inverse coverage factor $\bar{\kappa}$ is defined as in HHS84 to be the fraction of the total emission from the model which is picked up by the composite beam formed from the collection of pointing centers.
These quantities are shown in Table 1 for each galaxy in
the three lots.  For galaxies having $D_{H,e}$ less than the beam spacing
we are unable to determine $D_{H,e}$ and give only an upper limit.  Even
for well-resolved galaxies our uncertainty in this quantity remains
$1\arcmin\;$ or so.  For most galaxies $\bar{\kappa}$ is not much larger than
unity and the uncertainty in $\int S dV$ is comparable to the uncertainty
in an individual beam flux, $\sim$20\%.  
(In two cases the drop from central beam to adjacent points, due to observational error, is even more rapid than expected for a point source, so our algorithm gives total fluxes slightly smaller than the central beam flux.)
But IC 1613, because of its huge
extent, was mapped with beam spacings much larger than the beam-width
and so has much poorer coverage; the uncertainty in the flux for this
one galaxy is perhaps 50\%.

We also derive a maximum \ion{H}{1} extent $D_{H,max}$ from the outermost points at which
we claim to have reliably detected flux in the mainbeam for each
galaxy.  This was determined by careful inspection of the outermost
profiles after sidelobe reduction as described above.  In most cases the
velocity of the gas contributing to the sidelobe is sufficiently
different from that contributing to the mainbeam that there is no
dilemma.  In the two cases (UGC 12613 and IC 1613)
of much more extended galaxies mapped with
the circular feed we considered as well where significant quantities of
gas must be in order to contribute to sidelobes of points still further
out, reasoning that the gas must reach at least this far (and probably
somewhat further).  Once the outermost point with mainbeam emission was
identified, we assumed that the flux was emitted mainly into the beam
about one-half beam-width (1.\arcmin6) further in than the center of the
beam.  Since we generally integrated long enough on these outer points
to achieve an rms noise of $\sim$2 mJy and galaxy profiles are generally
narrower than 100 km $\rm s^{-1} \;$, the diameter $D_{H,max}$ corresponds to an
isophotal diameter at a limit of $\sim 1.2 \times 10^{19}  ( \frac{\Delta V_{50}} {100 \;{\rm km/s}}  )  ( \frac{{\rm rms}} {2\;{\rm mJy}} ) {\rm atoms} / {\rm cm}^2$.  The coefficient is
equal to $9.8 \times 10^4 \; {\rm M}_{\sun} / {\rm kpc}^2$.  The uncertainty
in $D_{H,max}$ is $1\arcmin\;$ or so.

In some cases we were able to determine $D_{H,e}$ but not $D_{H,max}$ (and less frequently vice versa).
However, in spite of the strong effect that tidal interactions can have on the outermost gas, the two radii are tightly correlated as shown in Fig. 1a.
Linear regression (the bisector of the regressions assuming all error in one, or all in the other, variable) gives $R_{H,max} = (1.63 \pm .17)  R_{H,e}^{(0.970 \pm .050)}$ where $r = \frac{d} {2} \sin D$ and $d$ is the distance to the galaxy in kpc.
While in principle we should use the direct linear regression of $\log R_{H,max}$ on $\log R_{H,e}$ (assuming all error in $\log R_{H,max}$) to estimate a missing value of $\log R_{H,max}$ from a measured $\log R_{H,e}$ (and the converse), in practice the relationship is sufficiently tight that neither direct regression differs significantly from the bisector regression, and so 
we use the regression given above to fill in missing values where appropriate. 
We also favor the use of the more nearly isophotal $R_{H,max}$ over $R_{H,e}$ in most applications.

\placefigure{radfig}

A number of these resolved dwarfs have distances determined from
primary distance indicators; the references are given in Table 3 below.  For
the other galaxies, all outside the Local Group, we infer the distance
from the galaxy's velocity with respect to the barycenter of the Local
Group (Yahil, Tammann \& Sandage \markcite{YTS77} 1977), using a Virgocentric infall
model that gives an asymptotic $H_o = 74$ km $\rm s^{-1}$/Mpc and a Local
Group deviation from Hubble flow of 273 km $\rm s^{-1} \;$ toward Virgo.  The assumed
distance to the Virgo Cluster is 19.04 Mpc (Rowan-Robinson \markcite{R88} 1988), and
we have taken the mean heliocentric redshift of the cluster to be 1135
km $\rm s^{-1} \;$ (HHS88).  Galaxies for which the solution is triple-valued have been
positioned at the distance most consistent with an application of the
blue Tully-Fisher relation or with their association with other galaxies
in a group.  With one exception (DDO 131), these distances are
consistent with those derived by Tully \markcite{T88} (1988).

In Figure 1b we compare the hydrogen and optical radii for all galaxies in this sample.
Four ``special case'' objects (discussed below) are singled out by special symbols, and upper limits  to the \ion{H}{1} radii are shown at half the formal limit as triangles with their apices pointing in the appropriate sense.  
The figure plots $R_{H,max} = \frac{d}{2} \sin D_{H,max}$ against $R_{25} = \frac{d}{2} \sin D_{25}$.  With the exception of DDO 154 and a couple of galaxies that are members of interacting systems, which we discussed in Hoffman et al. \markcite{HLSFLR93} (1993), the dwarfs fall along a well-defined correlation between the \ion{H}{1} and optical radii. 
Linear regression, excluding those special cases but including upper limits at half the formal value, gives

\begin{displaymath}
R_{H,max} = (2.67 \pm 0.19) R_{25}^{1.131 \pm 0.084} .
\end{displaymath}. 

\noindent
Within the uncertainties, this is consistent with the slightly shallower power law we get by
expanding this analysis to include dwarfs mapped by other observers and a corresponding sample of spirals, as we show in Paper II.

It is common practice in single beam observations to correct the
observed \ion{H}{1} flux for the finite size of the telescope beam, normally
by applying a correction factor derived from the beamwidth and the
optical diameter of the observed galaxy.  In Figure 2a we present a
scatter diagram of the ratio of total to central beam flux vs.
${D_{25}}^2$ for these Arecibo observations.
The correlation for these predominantly Sm and Im galaxies is quite
weak and dominated by the two largest galaxies; for the smaller objects there is essentially no correlation.  
Nevertheless, linear regression gives the ratio $R_F = (1.41 \pm 0.05) + 
(0.035 \pm 0.013) {D_{25}}^2$, where $D_{25}$ is measured in arcmin.  IC
1613, DDO 47, DDO 70 and DDO 154 were excluded from the regression since
the dual circular feed positioned on the center would receive
substantial sidelobe contributions to the measured flux.  The corresponding plot of $R_F$ vs. ${D_{H,e}}^2$ is given in Fig. 2b; not surprisingly, the correlation is somewhat better, but still not strong.  Linear regression gives $R_F = (1.25 \pm 0.06) + (0.023 \pm 0.004) {D_{H,e}}^2$.

\placefigure{FluxRat}

\subsection{Systemic velocities and profile widths}

Galaxy-wide systemic velocities and profile widths, intended to be
comparable to those obtained from much broader beams, are computed as
follows.  For the central profile, we found the points at 50\% of each
peak separately (or on each side of the peak in the case of
single-peaked profiles), measuring outward.  Uncertainties in these edge
velocities were determined following the discussion in Schneider et
al. \markcite{STMW90} (1990).  Then the outer 50\% edges of profiles obtained toward either
end of the major axis were determined similarly, along with their
uncertainties.  Finally, all corresponding edges were averaged, weighted
according to their respective uncertainties, giving global mean edge
velocities.  The average of these global mean edge velocities is taken
to be the heliocentric systemic velocity $V_{\sun}\;$; the difference is the
profile width $\Delta V_{50}$.  This procedure has the effect of
weighting the innermost 3 (or 5, for resolved galaxies) profiles about
equally; outer profiles, which have much less flux, invariably have
lower signal-to-noise and thus larger uncertainties, and contribute less
to the mean.  In the few cases of very well resolved galaxies with
rotation curves that rise appreciably between the outer edge of the
central beam and those beyond, the central profile has much more
gradually sloped edges than the next few further out, and accordingly
has larger uncertainty and lower weight.  In all cases $\Delta V_{50}$
calculated in this manner agrees well with that calculated from the
simple sum of all profiles acquired at the various points within the
galaxy.

\subsection{Rotation curves and dynamical masses}

The majority of our galaxies are too poorly resolved by the Arecibo beam
for us to say anything about the rotation curve or velocity field beyond
a determination of the direction of spin.  However, for 20 of the
galaxies we have enough resolution to determine rotation velocities for
at least two beams out from the center on each side of the major axis.
In no case is this sufficient resolution for us to explore in any detail
the rising, solid-body portion of the rotation curve, but we can examine
whether the outer parts of the rotation curves (generally resolved out of synthesis array maps) continue to rise, or become
flat as for most large spiral galaxies.

Galaxies at the low-mass end of our sample are presumed to be supported
as much by turbulent motions as by systematic rotation (Tully et al. \markcite{TBFGSvW78} 1978; Sargent, Sancisi \& Lo \markcite{SSL83} 1983; Skillman et al. \markcite{SBMW87} 1987; Lake \& Skillman \markcite{LS89} 1989; Lo, Sargent \& Young \markcite{LSY93} 1993).
Therefore it is of some importance to try to disentangle from our
spectra what part of the broadening of the feature is due to rotation
and what part to turbulent dispersion within the beam.  For a thin gas
disk rotating with no dispersion and a flat rotation curve, a spectrum
taken at some point along the major axis away from the galaxy's center
would be roughly triangular, with a sharp edge at the rotation velocity as shown in Fig. 3.
Rotation velocities that rise steeply across a major fraction of the
beamwidth will blunt that edge, especially if the gas density falls
steeply over the same length-scale.  (This is evident in 2 or 3 of our
most resolved and most rapidly rotating galaxies, DDO 154 and VCC 848 in
particular.)  Consequently we must be careful in applying the following
analysis
to the central beam spectra; but the outer spectra, where the rotation
curve appears to be flat or at most slowly rising in most cases, should
give reasonable estimates for the rotation velocity and reasonable upper
limits to the velocity dispersion averaged over the beam.

\placefigure{specfig}

The effect of velocity dispersion in the gas is to convolve the
roughly triangular rotation profile with a velocity spread function.  
As a first
approximation we will assume (as in Hoffman et al. \markcite{HLSFLR93} 1993) that the rotation profile is a perfectly
sharp-edged right triangle and the convolving function a Gaussian.  We
characterize the triangle by its width $W$ which is approximately equal
to $V_i \equiv V_{rot} \sin i$ if the beam is entirely to one side of the
kinematic center of the galaxy.  If $W \gg \sigma_z$, where $\sigma_z$ is the line-of-sight velocity dispersion, the convolution gives a
profile which has an edge velocity at 50\% of the profile peak
precisely equal to $V_i$ (assume that the systemic velocity of the galaxy has been subtracted out); the difference between the 20\% and 80\% edge velocities ${\Delta_{20}}^{80}$ can be found numerically to be about $1.20 \sigma_z$.  
In Hoffman et al. \markcite{HLSFLR93} (1993) we found that the ratios $\sigma_z / {\Delta_{20}}^{80}$ and $( V_{50} - V_i ) / V_{50}$ vary with the ratio of observables ${\Delta_{20}}^{80} / V_{50}$, following curves that are reasonably well represented by 

\begin{equation}
\frac{V_{50} - V_i} {V_{50}} = 1.11 x^2
\end{equation}

and

\begin{equation}
\frac{\sigma_z} {{\Delta_{20}}^{80}} = 0.42 \frac{e^{6x} - 1} {e^{6x} + 3.6} + 0.83,
\end{equation}

\noindent
where $x = {\Delta_{20}}^{80} / V_{50}$.
Measurements of the velocities at 20\%, 50\% and 80\% of peak flux
therefore give us approximations to $V_i$ and upper limits to
$\sigma_z$.  Note that the assumptions often made in determining
rotation curves from synthesis mapping, that the rotation velocity at
each point is either the intensity-weighted mean velocity or the center
velocity of a Gaussian fit to fluxes at that pixel, do not properly
account for the asymmetry due to beam-smearing and generally estimate a
rotation velocity that is too low.

Inclinations of dwarf galaxies are notoriously difficult to determine
(HHS88; Tully et al. \markcite{TBFGSvW78} 1978; Huchtmeier, Seiradakis \& Materne \markcite{HSM81} 1981; Sargent, Sancisi \& Lo \markcite{SSL83} 1983; Carignan, Beaulieu \& Freeman \markcite{CBF90} 1990 --- hereafter CBF; Lo, Sargent \& Young \markcite{LSY93} 1993).  
Our observations do not provide enough spatial
resolution to determine a kinematical inclination; nor, in most cases,
can we determine an \ion{H}{1} axis ratio precisely enough to be of help.
Therefore we are forced to resort to inclinations derived from optical
axis ratios in most cases.  For those dwarfs in which rotation dominates
($V_{rot} \gg \sigma$) it is probably not a bad assumption to suppose
that the stellar image mirrors the symmetry of the \ion{H}{1} disk.  Smaller
galaxies with weaker rotation ($V_{rot} \lesssim \sigma$) are not likely to
have such a correspondence.  The few that have been mapped with
synthesis telescopes (DDO 216, DDO 155, IC 1613, VCC 1816) give mixed
results:  DDO 155 in particular shows no correspondence between the
optical ``disk'' and \ion{H}{1} kinematics (CBF) while IC 1613 and VCC 1816
show reasonable agreement between those quantities.  Fortunately the
optically derived inclinations turn out to be near $45\arcdeg\;$ for all these
cases so that conservative corrections are applied.  Where synthesis
mapping exists and exhibits a rotation pattern ordered well enough to
allow a ring by ring analysis of the \ion{H}{1}
kinematics, we have preferred the inclination determined kinematically
from the outermost rings over that obtained by other means.

The dynamical mass has sometimes been estimated from the condition
for dynamical 
equilibrium in the plane of axisymmetry (Tully et al. \markcite{TBFGSvW78} 1978; Skillman et al. \markcite{SBMW87} 1987; Lake \& Skillman \markcite{LS89} 1989; Carignan, Beaulieu \& Freeman \markcite{CBF90} 1990; Lo, Sargent \& Young \markcite{LSY93} 1993):
 
\begin{displaymath}
\frac{{\rm G} M (r)} {r^2} = \frac{{V_{rot}}^2} {r} - {\sigma_z}^2 \frac{\partial} {\partial r} \ln \rho
\end{displaymath}

\noindent
where an isotropic velocity dispersion $\sigma$ and spherically symmetric
mass distribution have been assumed. 
For an isothermal sphere ($\rho \propto r^{-2}$ in the outskirts) this would give $( {V_{rot}}^2 + 2 {\sigma_z}^2 ) / r$ for the right-hand side.
However, a distribution of circular orbits of random orientation at radius $r$ would give an observed average line-of-sight velocity $\overline{{V_z}^2} = {V_{rot}}^2 / 3$, and the virial theorem applied to a uniform density (implied by the solid body rotation curves of the {\it optical} parts of many dwarf galaxies) sphere with constant and isotropic velocity dispersion (and negligible rotation) gives ${\rm G} M = 5 {\sigma_z}^2 r$.  As a compromise, we adopt

\begin{displaymath}
M = \frac{({V_{rot}}^2 + 3 {\sigma_z}^2 ) r} {\rm G} = 2.325 \times 10^5 {\rm M}_{\sun} \left ( \frac{{V_{rot}}^2 + 3 {\sigma_z}^2} {{\rm km}^2 / {\rm s}^2} \right ) \left ( \frac{r} {\rm kpc} \right ) .
\end{displaymath}

\noindent
Distances are determined as described in Sect. 4.2, above.

It is of interest to compare the total dynamical mass, determined as
described above, with the total luminous mass as deduced from the
optical and \ion{H}{1} flux received from the galaxy.  We take the mass in
stars to be $M_* = ( \frac{M} {L_B} ) L_B$, with 

\begin{displaymath}
\frac{M}{L_B} = 0.668 + 0.039 {( 10 - T )}^2
\end{displaymath}

\noindent
in solar units, $T$ being the morphological type index as given in de Vaucouleurs et al. \markcite{RC3} (1991 --- hereafter RC3).  The analytical expression is a fit to the models of Larson \& Tinsley \markcite{LT78} (1978); note that $\frac{M}{L_B}$ is model-dependent and not well-known (Arimoto \& Yoshii \markcite{AY86} 1986; Jablonka \& Arimoto \markcite{JA92} 1992).
We will discuss this point further in a later paper.

The hydrogen gas presumably has a nearly primordial admixture of helium, so that the total neutral monoatomic gas mass is $\frac{4} {3} M_{HI}$, where $M_{HI}$ is inferred from the observed \ion{H}{1} flux integral in the usual way (see Table 2).  
We do not know what fraction of the hydrogen is in molecular form and defer all discussion of that fraction to a later paper.

There is also ionized hydrogen in these galaxies, and the amount could be significant (Corbelli \& Salpeter \markcite{CS93a} 1993a, \markcite{CS93b} b; Maloney \markcite{M93} 1993; Salpeter \markcite{Sal93} 1993).
We can obtain a rough estimate of the amount as follows:
Corbelli \& Salpeter \markcite{CS93b} (1993) have shown that the H gas will be 50\% ionized by extragalactic UV at a face-on column density of $\sim 2 \times 10^{19}$ atoms ${\rm cm}^{-2}$.
The transition from mostly neutral to mostly ionized is quite sharp, and the exact amount of ionized gas inside that radius is model-dependent, but for a first approximation we might assume that surface density of ionized gas inside is uniform at about half the total H surface density at the truncation radius.
For the galaxies in Table 2 below, that produces estimates of \ion{H}{2} mass in the range 10-30\% of the measured \ion{H}{1} mass for most of the galaxies.  IC 1613, VCC 1816 and DDO 154 are exceptions, with \ion{H}{2} to \ion{H}{1} mass ratios of 0.38, 0.60 and 0.46 respectively.

\subsection{Results}

The basic data for all mapped galaxies are given in Table 1, with the
three samples merged into a single list ordered by R.A.  The table is organized with two rows per galaxy as follows:

\noindent
{\it Column (1):}  an identification number.

\noindent
{\it Column (2):}  up to 2 names are given for each galaxy.  Names
beginning with ``FS'' refer to the running entry numbers in the Leo group
catalog of Ferguson \& Sandage \markcite{FS90} (1990); ``VCC'' similarly refers to the
entries in the BST catalog of the Virgo Cluster.  Names starting with
``N,'' ``I,'' or ``U'' are drawn from NGC, IC, or UGC as usual.  Names like
``$9\arcdeg5$'' refer to the Virgo dwarf catalog of Reaves \markcite{R83} (1983), and GR 8
refers to Reaves \markcite{R56} (1956).  ``BTS'' galaxies are entries in the field dwarf
survey of BTS.  The galaxies from the surveys of van den Bergh (\markcite{vdB59}1959,
\markcite{vdB66}1966) are given the usual ``DDO'' notation.

\noindent
{\it Column (3):}  the galaxy's coordinates are given, with right
ascension (format hhmmss.s) in the upper row, declination (ddmmss) in
the lower row.  Both are from epoch 1950.

\noindent
{\it Column (4):}  the apparent $B$ magnitude and log of the diameter (in
units of 0.\arcmin1) at
isophote 25 mag/${\rm arcsec}^2$ are given.  No corrections for
extinction, either within the galaxy or the Milky Way, are applied.
Galaxies for which BST list only the estimated diameter (at $\sim 25.5$
mag/${\rm arcsec}^2$) have $D_{BST}$ transformed by the
correlation $\log D_{25} = 0.86 \log D_{BST} + 0.11$, derived in
HLHSW.  For the galaxies in FS we similarly obtain $\log D_{25} = 0.94 \log D + 0.15$ from the 9 Leo group galaxies that have both $D_{25}$ and $D_{FS}$ available; and for the BTS field survey galaxies
we find $\log D_{25} = 0.84 \log D_{BTS} + 0.22$, based on 49 galaxies
with both diameter measures available.  In all these transformation
relations, all $\log D$'s have $D$ measured in units of 0.\arcmin1.

\noindent
{\it Column (5):}  the morphological type of the galaxy is given in the
upper row, and a code for the source of the optical information is in
row 2.  These codes are explained in Table 3.

\noindent
{\it Column (6):}  the upper row contains a measure of the steepness of the profile edges, ${\Delta_{20}}^{80} \equiv \frac{\Delta V_{20} - \Delta V_{80}} {2}$, in units of km $\rm s^{-1} \;$.  
Here ${\Delta_{20}}^{80}$ is a weighted average for all spectra acquired for the galaxy, {\it not} a measure for a single beam position as described in section 4.4 above.  
The lower row contains the inverse coverage factor $\bar{\kappa}$ defined in section 4.2.

\noindent
{\it Column (7):}  the heliocentric systemic velocity and profile width
$\Delta V_{50}$, both in km $\rm s^{-1} \;$ and calculated as described in section
4.3 above, are given in the upper and lower rows, respectively.

\noindent
{\it Column (8):}  the upper and lower rows give the formal uncertainties in
the corresponding quantities of Col. (7).  Units are km $\rm s^{-1} \;$.

\noindent
{\it Column (9):}  the beam spacing used in mapping the galaxy and the
optical diameter $D_{opt} = 1.5 D_{25}$, in arcmin, are
given.

\noindent
{\it Column (10):}  the \ion{H}{1} diameters at $1/e$ times the central flux
and at the outermost point with detectable mainbeam flux appear in the
upper and lower rows, respectively.  These quantities are defined in
section 4.2 above.

\noindent
{\it Column (11):}  the upper and lower rows report the global galaxy
\ion{H}{1} flux integral, computed as described in section 4.2 above, in
units of Jy-km/s, and the representative rms noise (units mJy)
in the outermost few spectra from the map.

\placetable{alldwfs}

\section{Galaxies Resolved at Arecibo}

Table 2 gives some additional information for the 20 resolved galaxies.
The entries are ordered by R.A., as in Table 1.
Each galaxy is represented by two rows, organized as follows:

\noindent
{\it Column (1):}  upper row:  the galaxy name.  Lower row:  a reference to the galaxy's order of appearance in Table 1. 

\noindent
{\it Column (2):}  upper row:  the adopted distance to the galaxy, in Mpc.
Lower row:  the adopted inclination of the galaxy, in degrees.

\noindent
{\it Column (3):}  codes for the sources of the information in the
corresponding rows of Col. (2).  The codes are elaborated in Table 3.

\noindent
{\it Column (4):}  the inclination-corrected rotation velocity at the
outermost point on the rotation curve and the inferred galaxy-wide
average line-of-sight \ion{H}{1} velocity dispersion, both in km $\rm s^{-1} \;$.  See
section 4.4 above for details.

\noindent
{\it Column (5):} the radius, in kpc, of the outermost detected \ion{H}{1} (top row) and the \ion{H}{1} mass $M_{HI} = 2.356 \times 10^5 \; {\rm M}_{\sun}
\;d^2 \int S dV$ where $d$ is the distance to the galaxy in Mpc (bottom row).

\noindent
{\it Column (6):}  upper row:  the mass in stars defined as in section 4.4 above.
Lower row:  the total dynamical mass computed as described in section 4.4 above.  
Both are in units of of $10^8 \; {\rm M}_{\sun}$.

\noindent
{\it Column (7):}  the ratio of maximum \ion{H}{1} diameter to optical diameter, and the ratio of \ion{H}{1} diameter at $1/e$ of the peak flux to the optical isophotal diameter.

\noindent
{\it Column (8):}  upper row:  the blue luminosity $L_B$ in units of $10^8\;{\rm L}_{\sun}$.
Lower row:  the fraction of the observed \ion{H}{1} lying outside $D_{25}$, estimated from the ratio $D_{H,e} / D_{25}$ assuming a flattopped exponential \ion{H}{1} distribution.

\placetable{resnew}

\placetable{sources}

\placefigure{admaps}

\placefigure{lvmaps}

\placefigure{rotcurfig}

Fig. 4 shows contour maps of the total \ion{H}{1}
distribution for the 4 galaxies that we have mapped with sufficiently many beams.
Fig. 5  shows position-velocity contour maps for the resolved galaxies along with any others for which a clear sense of rotation was obtained.
Rotation curves for the resolved galaxies are shown in Fig. 6.  
Except
where otherwise noted, the contour maps use sidelobe-reduced spectra
throughout.  To produce the rotation curves, we deconvolved the effects of
dispersion from rotation as described in Sect. 4.4 above.  The rotation
velocity inferred from the two edges of the central beam, of course,
is an average over gas within the mainbeam at that position and should
be thought of as appropriate to a point $1\arcmin\;$ or so out from the center.
In essentially all cases except DDO 155, the bulk of the solid-body
rising rotation curve takes place within this central beam, and these
rotation curves are more nearly flat than solid-body.  Further details of the rotation curves are given in Table 4, as follows:

\noindent
{\it Column (1):}  top rows:  up to 2 names for the galaxy.  In the last row for each galaxy appears the ``global'' average velocity dispersion, a weighted average over all estimates obtained from individual beam positions.  The units are km $\rm s^{-1} \;$.

\noindent
{\it Column (2):}  top row:  a reference to the galaxy's order of appearance within Table 1.  Bottom row:  the maximum observed line-of-sight component of the rotation velocity, in km $\rm s^{-1} \;$.

\noindent
{\it Column (3):}  the beam's position relative to the center of the galaxy.  Numerical entries give the distance from the center in arcmin; the prefix (and suffix) give the direction.  ``E,'' ``N,'' etc. are the four ordinal directions, as usual.  A ``$+$'' or ``$-$'' is eastward (westward) along the major axis as determined either from the optical image or kinematically from preliminary mapping along the ordinal directions.  The suffix ``X'' implies mapping along the minor axis of the galaxy.

\noindent
{\it Column (4):}  the difference between the observed $V_{50}$ and the systemic velocity of the galaxy determined from the ensemble of $V_{50}$ measurements as described in section (c) above.

\noindent
{\it Column (5):}  the difference $V_{20} - V_{80}$ for the profile edge.

\noindent
{\it Column (6):}  the weight given each beam position, equal to $1 / {\sigma_{50}}^2$ where $\sigma_{50}$ is the uncertainty in $V_{50}$ for the edge, in km $\rm s^{-1} \;$.

\noindent
{\it Column (7):}  the velocity dispersion inferred from the analytical fits to the curves in Eq. (2).

\noindent
{\it Column (8):}  the line-of-sight component of the rotation velocity inferred from the analytical fits to the curves in Eq. (1).

\placetable{rotcurtab}

\section{Special Cases}

\subsection{Comparison to synthesis mapping}

A number of the dwarfs mapped by us have also been mapped with synthesis arrays:
The mappings are complementary in the sense that the Very Large Array or Westerbork provides finer resolution of the central parts of the galaxies while Arecibo offers greater sensitivity to the low density gas on the outskirts and, in most cases, better velocity resolution.  
In the best of cases, the synthesis maps allow a determination of the inclination of the \ion{H}{1} disk and detail of the inner parts of the rotation curve, and Arecibo extends the rotation curves by as much as a factor of 2 out from the galaxy's center.
We discuss the several best cases individually below.  
Three Virgo cluster galaxies, IC 3023, IC 3522 and UGC 7906, mapped at the VLA in its C and D arrays by Skillman \& Bothun \markcite{SB86} (1986) and Skillman et al. \markcite{SBMW87} (1987), are unresolved by Arecibo.
Table 5 reports comparisons of synthesis \ion{H}{1} diameters and fluxes with those obtained at Arecibo.  While all diameters resolved at Arecibo are substantially larger than the synthesis diameters, the gas on the outskirts is sufficiently low density that Arecibo fluxes exceed the synthesis fluxes by more than a few percent in only 2 of the 12 cases.
However, the synthesis maps and Arecibo maps do not agree so well on the shape of the outermost parts of the rotation curve:  of the 9 maps in common with sufficient spatial resolution at Arecibo for a rotation curve to be drawn, 4 are flat (or have opposite sides that have opposite trends) at Arecibo while the corresponding VLA maps are still rising at the outermost points on both sides;
3 are falling somewhat on both sides at the outermost points in the Arecibo maps while the VLA maps are flat; one (DDO 155, discussed below) is rising in the Arecibo map, falling in the VLA maps;
and only one rises on both sides in both maps.
The origin of these discrepancies is presumably the zero-spacing flux missing in those velocity channels of the VLA maps in which the emission is most extended.

\placetable{synthAO}

In the listing that follows, the numbers in parentheses following the galaxies' names are their reference numbers from Table 1:

\noindent
{\bf IC 1613 = DDO 8} (1) was mapped with the VLA hybrid C/D configuration by Lake \& Skillman \markcite{LS89} (1989 --- hereafter LaSk).  
The \ion{H}{1} extent in the VLA map is about 25\arcmin, with a total flux integral of 361 Jy-km $\rm s^{-1}$; the rotation curve rises linearly to the edge of the map.  
We find a total extent about a factor of 2 larger, $\sim$54\arcmin, with a total flux integral of 698 Jy-km $\rm s^{-1}$.  
The latter has a rather large uncertainty due to our undersampling of the \ion{H}{1} distribution; but it is not surprising that significant flux has been resolved out of the VLA image. 
Maps with larger beams have also been made, by Rots \markcite{R80} (1980), by Huchtmeier, Seiradakis \& Materne \markcite{HSM81} (1981) and by LaSk; these find flux integrals of 542, 290 and 481 Jy-km $\rm s^{-1}$, respectively. 
We find that the rotation curve becomes flat by $\sim 10\arcmin\;$ out from the center on the E side of the galaxy.  
The situation on the W side is murkier; at $-17.\arcmin1$ our algorithm finds no rotation, only a velocity dispersion twice as large as elsewhere in the galaxy.  The outermost profiles are too low signal-to-noise for our algorithm to apply.  
If we instead assumed a constant velocity dispersion $\sim$10 km $\rm s^{-1} \;$ to be used for all profiles, the deconvolution would give a rotation curve on both sides of the galaxy  rising gradually to $\sim$13 km $\rm s^{-1} \;$ and falling only at the last point on the W side.  Such behavior is often indicative of a warp, but our data on the outskirts of the galaxy are too sparse for any conclusion to be drawn in this case.

The inclination of IC 1613 is particularly difficult to estimate.
Optical estimates (reviewed in LaSk) range from $36\arcdeg\;$ to $50\arcdeg\;$ with some suggestions that the outer envelope becomes more edge-on.  LaSk find $i = 38\arcdeg\;$ from the \ion{H}{1} isophotes.
We find, from the Arecibo data, a best fit of our \ion{H}{1} profiles to a symmetric disk if the inclination is $20\arcdeg$.
It is important to resolve this controversy over the galaxy's inclination, because LaSk concluded that the MOND parameter deduced from the IC 1613 data with an inclination of $38\arcdeg\;$ is incompatible with MOND results for larger spiral galaxies, while Milgrom \markcite{Mi91} (1991) argues that an inclination of $20\arcdeg\;$ would bring IC 1613 into line.

\noindent
{\bf DDO 47 = UGC 3974} (4) was mapped at the VLA in its D array by Comte, Lequeux \& Viallefond (\markcite{CLV85} 1985 --- hereafter CLV; also Viallefond \markcite{V90} 1990).
The higher resolution VLA map shows a quite lumpy \ion{H}{1} distribution in contrast to the quite regular contours in our Arecibo map.
Both maps show a well-ordered velocity field; the VLA map, which has apparently resolved out the outer envelope of the \ion{H}{1} (note the diameters in Table 5), ends with the rotation curve still rising, while the Arecibo map suggests that the curve flattens out just beyond that radius.

\noindent
{\bf DDO 70 = Sextans B} (9) also shows a lumpy structure in the VLA D array map of CLV while the smoothing done by the large Arecibo beam produces much more regular contours.
In this case the VLA has resolved out significant flux; Arecibo detects 30\% more flux even though the diameter measured by Arecibo is only 30\% larger than that measured by the array.
The velocity field in the VLA map is also rather complicated, although it clearly rises all the way to the edge of the map; Arecibo also runs out of gas before the rotation curve turns flat.

\noindent
{\bf VCC 381 = $\bf{6\arcdeg18}$} (36) was mapped by Skillman et al. \markcite{SBMW87} (1987 -- hereafter SBMW) using the C array of the VLA.
The Arecibo map is barely resolved but suggests a diameter and flux both somewhat larger than  seen in the higher resolution map.
We see very little indication of an ordered velocity field in the Arecibo map and have not bothered to reproduce the major axis-velocity contour map here; at best we can only say that the northern end of the galaxy appears to have an average velocity about 4 km ${\rm s}^{-1} \;$ higher than the southern end.
The VLA map confirms the impression that the major axis is nearly N-S and shows a velocity field that rises roughly linearly with the northernmost isovelocity contour about 15 km ${\rm s}^{-1} \;$ higher than the southernmost.

\noindent
{\bf BTS 145 = CVn Dw A} (53) was mapped by LSY using the VLA in its C array.
As Table 5 indicates, significant diffuse flux has been resolved out of the VLA map (which has a rather short integration time).
The VLA and Arecibo maps both show an \ion{H}{1} distribution elongated more or less along the optically-defined major axis of the very low surface brightness image of the galaxy on the POSS.
However, that appearance is misleading; the Arecibo position-velocity maps in Fig. 5 indicates more ordered velocity structure along the optically-defined {\em minor} axis than along the major axis.
The kinematical axes as determined by Arecibo and the VLA agree; positions to the NE have higher velocity than positions to the SW.

\noindent
{\bf VCC 1816 = $\bf{14\arcdeg63}$} (55) was mapped by Helou \markcite{H84} (1984) with the C array of the VLA.
The structure of the galaxy in the VLA map is quite irregular and clumpy, and much diffuse emission has evidently been resolved out.
The velocity field is also quite irregular, although the VLA and Arecibo agree on a NS kinematical axis; when smoothed to the Arecibo spatial resolution, both velocity fields indicate a generally rising rotation curve.

\noindent
{\bf DDO 154 = UGC 8024} (60) was discussed in detail in Hoffman et al. \markcite{HLSFLR93} (1993 --- hereafter called HLSFLR).
After initial mapping at Arecibo by Krumm and Burstein \markcite{KB84} (1984), it was mapped by Carignan and Beaulieu (\markcite{CB89}1989, also Carignan and Freeman \markcite{CF88} 1988) at the VLA using the D array.  
The VLA image shows a quite regular disk with a modest warp evident at the SW end. 
As shown in Table 5, the considerable extent of this galaxy caused the VLA to resolve out significant flux; the Arecibo total flux and $D_{H,max}$ are both considerably larger than the corresponding quantities measured at the VLA.
In the VLA map, the rotation curve rises for the inner $4 \arcmin$ on either side, then flattens and appears to drop at the end.  
However, the drop may be due to a warp or to non-circular motion of the gas in the outer disk. 
The Arecibo map suggests a warp of $10\arcdeg \;$ or so, and the major axis position-velocity contour map presented in HLSFLR suggests that the rotation curve, after dropping by $\sim 5$ km $\rm s^{-1} \;$ at a radius of $\sim 7 \arcmin$, holds constant at that lower level to the galaxy's edge. 
In HLSFLR we modeled the \ion{H}{1} distribution with a series of inclined thin rings with circular gas orbits and a radius-dependent local velocity dispersion; the best-fit model has a rotation curve which does continue to decline beyond the last point in the VLA map, but which never falls as rapidly as a Keplerian disk.

\noindent
{\bf GR 8 = UGC 8091 = DDO 155} (62) was mapped with the VLA C array by Carignan, Beaulieu \& Freeman \markcite{CBF90} (1990 --- hereafter CBF) and by LSY.  Although it is projected onto the outskirts of the Virgo cluster, the galaxy is apparently a member of the Local Group at a distance of about 1 Mpc (Hodge \markcite{H74} 1974, de Vaucouleurs \& Moss \markcite{dVM83} 1983; Aparicio, Garcia-Pelayo \& Moles \markcite{AGM88} 1988; see discussion in LSY) and has the lowest mass of all our ``resolved'' galaxies.  
The VLA images find an \ion{H}{1} extent of 3.\arcmin8 and an \ion{H}{1} flux of 7.2 Jy-km $\rm s^{-1} \;$.  Arecibo only just resolves the \ion{H}{1} extent and finds it to be consistent with the VLA extent, and our total \ion{H}{1} flux integral is 8.6 Jy-km $\rm s^{-1} \;$, only a bit larger than that found at the VLA; there is apparently no significant diffuse envelope resolved out of the VLA image.  
The rotation curve published by CBF rises to a maximum $\sim$0.\arcmin4 from the center, then falls in Keplerian fashion, but this does not seem consistent with the velocity contour maps shown in that paper; in any case it is not clear that it is correct to describe the gas kinematics as rotation.  
Our best estimate of the ordered and turbulent motions from the \ion{H}{1} profiles averaged over the Arecibo beam is that the central portion, to a radius of 1\arcmin or so, has essentially no ordered motion and a velocity dispersion $\sim$15 km $\rm s^{-1} \;$; but at the N edge we find a bulk velocity 10 km $\rm s^{-1} \;$ higher than the galaxy's systemic velocity while the E edge has a bulk velocity 10 km $\rm s^{-1} \;$ lower than systemic.  
Within the uncertainties this would be consistent with a generally rising rotation curve along a NW-SE axis, more or less in line with the velocity contour maps in CBF.

\noindent
{\bf DDO 187 = UGC 9128} (67)  was mapped by LSY using the C array of the VLA.
In the VLA map, the \ion{H}{1} distribution is reasonably smooth, in contrast to that for most of the other dwarfs, but not well aligned with the optical image; but the velocity field is quite irregular, although a general NS kinematical axis can be discerned.
The Arecibo map confirms the NS kinematics (the ``major axis'' of Fig. 5 is in fact the NS axis) but hints at a flattening (or declining) of the rotation curve at both ends.
Evidently this part of the rotation curve has been resolved out of the VLA image.

\noindent
{\bf DDO 216 = UGC 12613} (70) was mapped with the VLA C array by LSY.  
The image has a major diameter of 6.\arcmin1 and is distinctly asymmetrical, the pronounced \ion{H}{1} density peak being displaced by $\sim 1\arcmin\;$ from the center of the outer \ion{H}{1} isophotes.  
We have a diameter $\sim 8\arcmin\;$ and a flux integral $\sim$25\% larger than the VLA (LSY), so some diffuse \ion{H}{1} has apparently been resolved out of the VLA image.  
The asymmetry can be seen to some extent in the Arecibo map, but more surprising is the fact that the kinematical center of the galaxy is apparently displaced by $\sim$1.\arcmin9 in the opposite direction (as indicated by a classic double horned profile obtained at that location while the optical center produces a single-horned profile reminiscent of the profile we would expect to obtain from a beam pointed 1.\arcmin9 from the center of a typical spiral galaxy).  
The offset kinematical center is visible in the VLA velocity field too, which indicates roughly solid body rotation; the Arecibo map shows that the rotation curve becomes flat about 4.\arcmin5 from the kinematical center ($\sim$2.\arcmin6 from the optical center) toward the SE; on the NW side the rotation curve apparently falls after reaching a peak $\sim 2\arcmin \;$ from the kinematical center.  
This is most likely due to a warp or other break from circular motion, but our map is too sparsely sampled around the outskirts for us to be sure.

For distance-dependent quantities, we have assumed the Hoessel et al. \markcite{HAMSD90} (1990) distance of 1.7 Mpc for DDO 216 throughout.
There is some controversy about that distance, however; in particular, Aparicio \markcite{A94} (1994) obtains a much smaller distance of 0.95 Mpc.

\subsection{Interacting pairs}

In addition to NGC 4532/DDO 137 and NGC 4694 / VCC 2062, which were discussed  in HLSFLR, the galaxy samples given in Table 1 include two close, probably interacting pairs of galaxies:  FS 35/36 and $13\arcdeg118 / 13\arcdeg118$a.  

\noindent
{\bf FS 35/36} (13 \& 14):  This pair, Sm and ImIV, was identified by Ferguson and Sandage \markcite{FS90} (1990) on du Pont plates of a portion of the Leo complex of groups.  
The two are separated by 2.\arcmin7 and have a velocity difference of 39 km $\rm s^{-1} \;$ with profile widths of 74 and 25 km $\rm s^{-1}$.  
Ordinarily we could expect the profiles to be confused, but evidently we are fortunate; FS 36 has its peak \ion{H}{1} density $\sim 2\arcmin\;$ W of the center of its optical image, while FS 35's \ion{H}{1} peak lies $\sim 3\arcmin\;$ N of its optical center.  
Otherwise the \ion{H}{1} distribution of FS 36 is not atypical: a narrow Gaussian lump only just resolved by the Arecibo beam; there is some indication of elongation in the EW direction.  
FS 35, however, is not at all typical of Sm galaxies with profile widths as wide as 74 km $\rm s^{-1}$.  
On the southern end, the spectra are at least square-topped if not double-horned, as if they had been obtained near the center of a rotating disk, but all other points around the galaxy show Gaussian (or triangular) profiles with no clear evidence of rotation.  
Evidently the tidal interaction has severely disrupted the dynamics of the gas in the larger galaxy if not in both.

\noindent
{\bf NGC 4532 / DDO 137} (not included in Table 1):  This pair and its extended envelope of \ion{H}{1} were discussed in detail in HLSFLR and Hoffman et al. \markcite{HSLR92} (1992).
The Arecibo map shows asymmetric, distorted disks centered on each of the two galaxies with an enormous diffuse cloud having peculiar kinematics surrounding the pair.
There is also a long plume reminiscent of tidal features extending out from DDO 137, and a couple of denser cloudlets with \ion{H}{1} mass around $10^8 \; {\rm M}_{\sun}$ are evident within the cloud.
Subsequent synthesis array mapping, to be reported later, confirms the existence of these coudlets. 

\noindent
{\bf NGC 4694 / VCC 2062} (58 \& 59):  Our Arecibo data for this pair were discussed in HLSFLR.
VCC 2062, an Im galaxy as shown in HLSFLR, lies about 3.\arcmin8 W of NGC 4694, a $B_{T} = 12.19$ amorphous galaxy.
We presented a sparsely sampled spatial \ion{H}{1} map of this pair, which was also was mapped with the VLA D array by Cayatte et al. \markcite{CGBK90} (1990) and at Westerbork by van Driel and van Woerden \markcite{vDvW89} (1989).  
The synthesis maps both show a small cloud of gas in the center of the larger galaxy and a diffuse but more massive cloud whose highest density peak coincides with VCC 2062.  
All three maps show a linear gradient in velocity from the velocity of NGC 4694 to about 100 km $\rm s^{-1} \;$ less at the W edge of the cloud.
None of the maps found any indication of rotation for NGC 4694 itself.
The \ion{H}{1} data in combination with CCD data lead HLSFLR to the conclusion that
the VCC 2062 cloud of gas and stars is tidal in origin, with the dense stellar knot at the \ion{H}{1} peak resulting from a more recent star formation burst.

\noindent
$\bf{13\arcdeg118 / 13\arcdeg118a}$ (63 \& 64):  $13\arcdeg118$ (= CGCG 071-090 = MCG +02-33-043) was catalogued by Reaves \markcite{R83} (1983) from a IIIaJ Palomar Schmidt plate.  
It lies just E of the BST survey area.  To the N, E and W the \ion{H}{1} distribution appears typical:  Gaussian profiles, no clear indication of rotation, the \ion{H}{1} extent barely resolved by the Arecibo beam.  
But to the S the spectra acquire a component 44 km $\rm s^{-1} \;$ higher in velocity, indicating a separate cloud of \ion{H}{1} about $3\arcmin\;$ S of the optical center of $13\arcdeg118$.  
The blue POSS print was inspected after the \ion{H}{1} mapping was complete, and a small patch, $\sim$0.\arcmin5 in diameter and irregular in outline, was found at the indicated position.

\section{Discussion}

As we show in Figure 1b and will further document in Paper II, the vast majority of galaxies that have been mapped in \ion{H}{1} have \ion{H}{1} envelopes that do not exceed 5 times the optical extent.
Of the galaxies reported here, only DDO 154 has
$R_{max} \equiv D_{H,max} / D_{25} > 14$, and we know of only one other galaxy, \ion{H}{1} 1225+01 (Giovanelli, Williams \& Haynes \markcite{GWH91} 1991) that has larger $R_{max}$. 
Implications of this result are discussed in Hoffman et al. \markcite{HLSFLR93} (1993) and in Paper II.

Of the 19 rotation curves displayed in Fig. 6, only one (UGC 1281) is clearly flat on both sides of the major axis.
The other cases slightly favor rising on both sides as far out as the gas can be traced (8 cases) over falling on both sides at the outermost point (5 cases), and there are 5 cases in which the two sides give conflicting results.
We have no
convincing examples of Keplerian fall-off; the few cases of apparently
descending rotation curves at the outermost edges are most likely due to
warps or other non-circular gas motions although we do not have sufficient detail in most cases to be sure.  
The situation is very similar to that of spiral galaxies (see Persic, Salucci \& Stel \markcite{PSS95} 1995 and the review by Ashman \markcite{A92} 1992),
where rotation curves rise in solid-body fashion through some portion of
the optical disk, then remain flat (or at most slightly falling) as far as they can be traced; only
the total mass is much less in dwarfs, so the \ion{H}{1} column density has
become too low to support star formation at the edge of the dark matter
core.  The higher mass of spiral galaxies drives the central \ion{H}{1}
density higher, allowing star formation farther out into the $\frac{1} {r^2}$ halo.
In a few of the most centrally concentrated spiral galaxies the rotation curves decline on both sides of the major axis (Casertano \& van Gorkom \markcite{CvG91} 1991); however, Persic, Salucci \& Stel \markcite{PSS95} (1995) argue that this is strictly determined by the total luminosity of the galaxy, with the least luminous spirals exhibiting rotation curves that rise to the edge of the optical disk and the most luminous having disks that decline (but not so rapidly as a Keplerian fall-off).
We have explored whether either trend is apparent among the dwarf irregulars also.
The three smallest systems all exhibit curves that rise as far out as they can be traced, but there is no significant evidence that the least concentrated systems are more likely to have rising curves.  
None of our galaxies are luminous enough to fall among those that Persic, Salucci \& Stel \markcite{PSS95} (1995) cite for systematically declining rotation curves.

The panels of Fig. 7 explore the correlations among the various size indicators:  $ \log ( L_B / {\rm L}_{\sun} ), \; \log ( M_H / {\rm M}_{\sun} ) ,\; \log ( V_{c} / {\rm km} \;{\rm s}^{-1} ) $ and $\log ( R_{gm} / {\rm kpc} )$, where $R_{gm}$ is the geometric mean of $R_{H,max}$ and $R_{opt} \equiv 1.5\,R_{25}$, used so as to average over instrumental uncertainties in the measurement of $R_{H,max}$ and the intrinsic unreliability of $R_{25}$ as a measure of the true size of the dark matter halo of the galaxy.  
The dynamical velocity is taken to be $V_{c}^2 \equiv {V_{rot}}^2 + 3 {\sigma_{z}}^2$ as above.
We also consider the indicative dynamical mass $M_{dyn} \equiv V_{c}^2 R_{gm} / \rm{G}$.
Since there are measurement uncertainties in all variables, we use the bisector of the direct and inverse regressions (Isobe et al. \markcite{IFAB90} 1990).

\placefigure{correls}

The resulting correlations of $R_{gm}$, $V_c$, $M_H$, and $M_{dyn}$ against $L_B$, along with $V_c$ vs. $R_{gm}$ and $M_H$ vs. $M_{dyn}$, are plotted in Fig. 7, with $R_{H,max}$ vs. $R_{25}$ having appeared earlier in Fig. 1b.
In these figures, Sdm, Sm and Im galaxies appear as exes, while
the few BCD galaxies in our sample are indicated with open squares.
Asterisks are reserved for those that are in binary (possibly interacting) systems, which are omitted from all regressions.
In these cases we have done our best to identify that emission which can be clearly associated with each individual member of the pair and then determined $V_c$ and $R_{H,max}$ as if that emission were from an isolated galaxy.
Four special cases are shown with filled symbols and also omitted from all regressions:  DDO 154 with a circle, VCC 2062 with a diamond, DDO 137 with a triangle, and the NGC 4532 / DDO 137 complex as a whole (using the extreme \ion{H}{1} velocities observed anywhere in the map to determine $V_c$, defining $R_{H,max}$ for the \ion{H}{1} cloud as a whole, and using the sum of the optical radii of the two galaxies for $R_{opt}$) with a hexagon. 
The solid line in each plot is the best fit (bisector of direct and inverse least-squares fits).
We have also investigated whether the galaxies are segregated in these plots according to whether the rotation curve at the outermost observed point is still rising, has turned flat, or has begun to fall.
Within the limits of small number statistics, the slope of the rotation curve produces no discernable effect on any of the correlations.

The results of these correlations are as follows:

\begin{displaymath}
R_{gm} = (9.43 \pm .63) ({L_B} / {10^9 {\rm L}_{\sun}} )^{0.431 \pm .035}\; {\rm kpc}
\end{displaymath}

\begin{displaymath}
V_c = (84.8 \pm 6.9) ({L_B} / {10^9 {\rm L}_{\sun}} )^{0.384 \pm .047} \;{\rm km} \;{\rm s}^{-1}
\end{displaymath}

\begin{displaymath}
M_H = (7.84 \pm .98) ({L_B} / {10^9 {\rm L}_{\sun}} )^{0.888 \pm .075} \;{\rm M}_{\sun}
\end{displaymath}

\begin{displaymath}
M_{dyn} = (1.04 \pm .15) ({L_B} / {10^9 {\rm L}_{\sun}} )^{0.900 \pm .078} \;{\rm M}_{\sun}
\end{displaymath}

\begin{displaymath}
V_c = (72.7 \pm 4.5) ({R_{gm}} / {8.042\;{\rm kpc}} )^{0.863 \pm .08} \;{\rm km\;s}^{-1}
\end{displaymath}

\begin{displaymath}
M_H = (6.42 \pm .85) \times 10^8 ({M_{dyn}} / {1.212 \times 10^{10} {\rm M}_{\sun}} )^{0.965 \pm .071} \;{\rm M}_{\sun} .
\end{displaymath}

\noindent
In Paper II we will display similar correlations for this sample combined with all other available mapped dwarfs and a sample of mapped spirals spanning the same range of redshifts.
Within the rather substantial uncertainties, the correlations above are all consistent with those for the combined samples; however, the dwarf + spiral sample has slightly smaller powers of $L_B$ in each case ($0.382 \pm .013$, $0.276 \pm .014$, $0.759 \pm .033$, and $0.859 \pm .027$ for the four regressions in the order given above).
The most striking difference between this sample and that for the spirals alone is the much larger scatter in all correlations for dwarfs, and in most cases any offset between the correlations for dwarfs alone and for spirals alone is swamped by that scatter.
There is, however, an offset of a significant number of mapped dwarfs from the literature from our data and from mapped spirals in the $\log V_c$ vs. $\log L_B$ plane, those dwarfs forming a sequence parallel to the rest at lower luminosity for given rotation speed.
This is consistent with some authors (e.g. van Zee, Haynes \& Giovanelli \markcite{vZHG95} 1995) having selected specifically low surface brightness objects for mapping.
We will discuss this point further in Paper II.

\section{Summary}

\ion{H}{1} mapping of dwarf galaxies potentially answers a number of questions about the structure and dynamics of galaxies in general and about the process of galaxy formation.
Here we have presented Arecibo mapping of three samples totalling 70 objects.
These include a sample of galaxies from the Virgo Cluster Catalog (BST), another from the field dwarf survey of BTS, and a third intended to complete a distance-limited catalog of dwarfs with Arecibo mapping.
For these, Table 1 reports \ion{H}{1} masses and diameters, systemic velocities and profile widths.

Approximately a quarter of our sample is sufficiently resolved by the Arecibo beam for us to derive a rotation curve.
For each of these 20 objects, Table 2 reports further information:  a deconvolution into ordered rotation and random motion contributions to the profile width, an ``isophotal'' \ion{H}{1} diameter representing the outermost points at which we claim to detect emission, and a dynamical mass.
The rotation curves themselves appear in Table 3 and Figures 5-6.

Twelve galaxies from our sample have published synthesis array mappings, from either the Very Large Array or Westerbork.
We have made a detailed comparison of the results, in general finding (as expected) that Arecibo diameters are larger even though the total flux measured by Arecibo is more than a few percent larger in only two cases.
The outer parts of Arecibo rotation curves generally have flattened or begun to decline while the synthesis curves are generally still rising at the outermost point; as a general characterization we would say that dwarf rotation curves rise in more or less solid-body fashion to the edge of the optical disk and then remain flat as far as the \ion{H}{1} can be traced, with the outer edges in most cases corrupted by non-circular motions (warps or turbulence).

Very few of the \ion{H}{1} envelopes exceed the diameter of the stellar disk by more than a factor of 5; only DDO 154 (and \ion{H}{1} 1225+0146, mapped by Giovanelli, Williams \& Haynes \markcite{GWH91} 1991 and Chengalur, Giovanelli \& Haynes \markcite{CGH95} 1995) have \ion{H}{1} disks more than 10 times as extended as their stellar components.

Correlations among four variables --- luminosity, radius, \*(HI mass, and rotation velocity --- and a dynamical mass, derived from the rotation velocity and radius, were presented.  
While the mean relations are consistent with those found for the {\it Virgo Cluster Catalog} as a whole (Hoffman, Helou \& Salpeter \markcite{HHS88} 1988) and with the larger sample of mapped dwarfs and spirals discussed in Paper II, the scatter is large; this probably reflects measurement uncertainties, distance errors and very uncertain inclinations in addition to significant intrinsic scatter (especially in the optical luminosity).

In future papers we shall discuss the use of these dwarf galaxy mappings to test the Modified Newtonian Dynamics (MOND) of Milgrom (\markcite{Mi83a}1983a, \markcite{Mi83b}b, \markcite{Mi83c}c) and the upper limits to fermionic dark matter (Tremaine \& Gunn \markcite{TG79} 1979; Sciama \markcite{Sc90} 1990; Madsen \markcite{Ma91} 1991; Ralston \& Smith \markcite{RS91} 1991; Barnes \markcite{B93} 1993; Takahara, Komai \& Yamakawa \markcite{TKY93} 1993) that these data imply.

\acknowledgments
We acknowledge fruitful discussions and correspondence with G. Bothun, J. Dickey, R. Giovanelli, B. M. Lewis,  K. Y. Lo, M. Roberts and E. Skillman.  We thank C. Carignan for providing us with a copy of the VLA data for DDO 154, and K. Y. Lo and F. Viallefond for sharing results with us prior to publication.  T. Groff assisted with some of the observations of DDO 154.  The Arecibo staff was, as always, hospitable, attentive and helpful throughout these observations.  This work was supported in part by US National Science Foundation grants AST 84-06392, AST-8713394, AST-9015181 and AST-9316213 at Lafayette College and by AST 84-15162, AST-8714475 and AST-9119475 at Cornell, and in part by the National Astronomy and Ionosphere Center which is operated by Cornell University for the National Science Foundation.

\clearpage

\begin{figure}
\caption{
\label{radfig}
(a) Maximum hydrogen radius $R_{H,max}$ vs. hydrogen radius at 1/$e$, $r_e$, for sample galaxies that have both measured at Arecibo.  (b)  Maximum hydrogen radius $R_{H,max}$ vs. optical radius $R_{25}$, both in kpc and scaled by the corresponding radius appropriate to a galaxy with $L_B = 10^9$ solar luminosities, for our sample.  
Sm and Im galaxies are represented by exes, BCD by open squares.
Four special cases are identified with special symbols:  DDO 154 by a circle, DDO 137 by a triangle, the NGC 4532 / DDO 137 complex by a hexagon, and VCC 2062 by a diamond.  
Upper limits to $R_{H,max}$ are indicated at half the formal value by open triangles.  
The solid line is the bisector of the two standard linear regressions ($y$ vs. $x$ and $x$ vs. $y$).}
\end{figure}

\begin{figure}
\caption{
\label{FluxRat}
(1) The ratio of the total flux integrated over all beam positions to that in the central beam plotted against the optical surface area $\propto {D_{25}}^2$ (in units of arcmin).  All dwarfs mapped by us at Arecibo, except IC 1613, DDO 47, DDO 70 and DDO 154, are shown, as discussed in Sect. 4.2.
(b) The ratio of total flux to central beam flux plotted against ${D_{H,e}}^2$, $\propto$ area of the HI disk.}
\end{figure}

\begin{figure}
\caption{
\label{specfig}
Representative Arecibo neutral hydrogen spectra from the center of IC 3356 and from points 1\farcm9 on either side of the center along the major axis of the galaxy.
The steep sides of the off-center points indicate approximately the line-of-sight velocity of gas in the galaxy's disk at the point where the line-of-sight of the center of the Arecibo beam falls tangent to the circular orbit of the gas.
See text for further discussion.}
\end{figure}

\begin{figure}
\caption{
\label{admaps}
Contour maps of total neutral hydrogen emission in four well-resolved dwarfs.
For IC 1613, the contour levels range from 0.2 to 14.2 in steps of 2.0 Jy-km $\rm s^{-1} \;$ per beam; for DDO 47, from 0.5 to 16.5 in steps of 2.0 Jy-km $\rm s^{-1} \;$ per beam; for DDO 70, from 1.0 to 17.0 in steps of 2.0 Jy-km $\rm s^{-1} \;$ per beam; and for DDO 216, from 0.5 to 10.5 in steps of 2.0 Jy-km $\rm s^{-1} \;$ per beam.
Arecibo beam positions used in producing the maps are indicated by small dots.}
\end{figure}

\begin{figure}
\caption{
\label{lvmaps}
Position-velocity contour maps for various dwarfs.
The legend on each indicates the contour spacing:  ``LOG'' indicates that the contours are logarithmic, with the numerical values coding the log of the minimum flux in mJy and the spacing in square brackets giving the interval in the log; otherwise the minimum contour is given in mJy with the interval in square brackets also in mJy.
The outermost contour, typically at three times the rms flux in the outermost spectra, is shown with a dashed curve. }
\end{figure}

\clearpage

\begin{figure}
\caption{
\label{rotcurfig}
Rotation curves for resolved dwarfs.
Open symbols represent the observed half-power velocity, $| V_{50} - V_{sys} |$, with triangles and circles for the opposite sides of the axis (major axis unless indicated otherwise).
Each is plotted at the observed beam position, and the formal uncertainty, determined by the steepness of the sides and the signal-to-noise ratio in the spectrum at each point, is indicated by the error bars.
The corresponding solid symbols give the rotation speed $V_i$, inferred as in Sect. 4.4, where it can be calculated; no correction for galaxy inclination has been applied.
The vertical arrow in each figure gives the optical radius at the 25 mag/${\rm arcsec}^2$ isophote.}
\end{figure}

\begin{figure}
\caption{
\label{correls}
Correlations among size and dynamical variables.
Symbols are chosen as in Fig. 1.
The four special cases represented by solid symbols are shown for illustration only and are not included in the correlations.
In each panel, the solid line is the bisector of the two ordinary least-squares regressions.
(a)  Geometric mean of optical and hydrogen radii in kpc vs. blue luminosity in solar units.
(b)  Tully-Fisher diagram of dynamical velocity in km $\rm s^{-1} \;$, as discussed in the text, vs. blue luminosity in solar units.
(c)  Hydrogen mass vs. blue luminosity, both in solar units.
(d)  Indicative dynamical mass vs. blue luminosity, both in solar units.
(e)  Dynamical velocity profile width in km $\rm s^{-1} \;$ vs. radius in kpc.
(f)  Hydrogen mass vs. indicative dynamical mass, both in solar units. }
\end{figure}

\clearpage

\begin{table}
\dummytable
\label{alldwfs}
\end{table}

\begin{table}
\dummytable
\label{resnew}
\end{table}

\begin{table}
\dummytable
\label{sources}
\end{table}

\begin{table}
\dummytable
\label{rotcurtab}
\end{table}

\begin{table}
\dummytable
\label{synthAO}
\end{table}

\end{document}


\begin{deluxetable}{r l r r c r r c c r r}
\tablenum{1}
\tablewidth{0pt}
\tablecaption{Mapped Dwarfs}
\tablehead{
\colhead{\#} & \colhead{Name} & \colhead{R.A.} & \colhead{$B_T$} & \colhead{Type} & \colhead{${\Delta_{20}}^{80}$} & \colhead{$V_{\sun}$} & \colhead{$\sigma$} & \colhead{space} & \colhead{$D_{H,e}$} & \colhead{$\int SdV$} \\
& \colhead{Name} & \colhead{Dec.} & \colhead{$\log {D_{25}}$} & \colhead{source} & \colhead{$\bar{\kappa}$} & \colhead{$\Delta V_{50}$} & \colhead{$\sigma$} & \colhead{$D_{opt}$} & \colhead{$D_{H,max}$} & \colhead{rms} \\
\colhead{(1)} & \colhead{(2)} & \colhead{(3)} & \colhead{(4)} & \colhead{(5)} & \colhead{(6)} & \colhead{(7)} & \colhead{(8)} & \colhead{(9)} & \colhead{(10)} & \colhead{(11)} }
\startdata
1 & IC 1613 & 10219.7 & 9.99 & ImV & 9 & $-233$ & 1 & $5.\arcmin7$ & $24\arcmin$ & 698 \nl
 & DDO 8 & 15156 & 2.08 & LaSk & 4.9 & 27 & 1 & $18\arcmin$ & $\sim$54\arcmin & \nl
 & & & & & & & & & & \nl
2 & UGC 685 & 10443.0 & 14.5 & SAm & 18 & 157 & 1 & $1.\arcmin9$ & $2.\arcmin1$ & 11.8 \nl
 &  & 162501 & 1.29 & UGC & 1.1 & 74 & 3 & $2.\arcmin9$ & $\sim6\arcmin$ & 1.1 \nl
 & & & & & & & & & & \nl
3 & UGC 1281 & 14638.8 & 12.6 & Sdm & 10 & 157 & 2 & $1.\arcmin9$ & $6.\arcmin4$ & 38.9 \nl
 &  & 322033 & 1.64 & T & 1.3 & 123 & 4 & $6.\arcmin5$ & $\sim7\arcmin$ & 0.9 \nl
 & & & & & & & & & & \nl
4 & DDO 47 & 73903.0 & 13.54 & ImV & 15 & 270 & 1 & $1.\arcmin3$ & $7.\arcmin0$ & 67.0 \nl
 & UGC 3974 & 165506 & 1.66 & T & 1.5 & 55 & 1 & $6.\arcmin9$ & $\sim12\arcmin$ & 5.9 \nl
 & & & & & & & & & & \nl
5 & UGC 4115 & 75412.8 & 14.29 & ImIV-V & 15 & 341 & 4 & $1.\arcmin9$ & $< 1.\arcmin9$ & 21.4 \nl
 &  & 143120 & 1.30 & T & 1.2 & 95 & 8 & $3.\arcmin0$ & $\sim6\arcmin$ & 1.2 \nl
 & & & & & & & & & & \nl
6 & BTS 5 & 93454.0 & 16.0 & Im & 16 & 1597 & 3 & $1.\arcmin9$ & $2.\arcmin7$ & 3.2 \nl
 &  & 274736 & 0.91 & BTS & 1.0 & 41 & 6 & $1.\arcmin2$ & $\sim$4\arcmin & 1.1 \nl
 & & & & & & & & & & \nl
7 & DDO 64 & 94727.5 & 14.3 & ImIII & 15 & 521 & 1 & $1.\arcmin9$ & $5.\arcmin8$ & 18.5 \nl
 & UGC 5272 & 314318 & 1.34 & BTS & 1.3 & 84 & 2 & $3.\arcmin3$ & $\sim5\arcmin$ & 1.7 \nl
 & & & & & & & & & & \nl
8 & DDO 68 & 95353.2 & 14.8 & Im/BCD & 13 & 506 & 2 & $1.\arcmin9$ & $4.\arcmin1$ & 35.3 \nl
 & UGC 5340 & 290336 & 1.41 & BTS & 1.2 & 93 & 4 & $3.\arcmin9$ & $\sim8\arcmin$ & 1.2 \nl
 & & & & & & & & & & \nl
9 & DDO 70 & 95723.4 & 11.86 & ImIV-V & 14 & 304 & 1 & $1.\arcmin3$ & $7.\arcmin8$ & 102.4 \nl
 & Sextans B & 53418 & 1.66 & T & 1.4 & 40 & 2 & $6.\arcmin9$ & $\sim13\arcmin$ & 5.1 \nl
 & & & & & & & & & & \nl
10 & UGC 5456 & 100439.6 & 13.5 & pec & 29 & 540 & 2 & $1.\arcmin9$ & $< 1.\arcmin9$ & 6.4 \nl
 &  & 103634 & 1.28 & UGC & 1.2 & 62 & 3 & $2.\arcmin9$ & $\sim4\arcmin$ & 1.0 \nl
 & & & & & & & & & & \nl
11 & FS 3 & 103817.5 & 15.6 & SmIV & 11 & 1013 & 1 & $1.\arcmin9$ & $< 1.\arcmin9$ & 1.8 \nl
 & UGC 5812 & 124415 & 1.00 & FS & 1.2 & 55 & 2 & $1.\arcmin5$ & $< 4\arcmin$ & 1.9 \nl
 & & & & & & & & & & \nl
12 & FS 32 & 104617.9 & 15.2 & BCD & 17 & 1324 & 1 & $1.\arcmin9$ & $2.\arcmin2$ & 5.0 \nl
 &  & 122716 & 0.84 & FS & 1.1 & 125 & 3 & $1.\arcmin0$ & $\lesssim4\arcmin$ & 2.7 \nl
 & & & & & & & & & & \nl
13 & FS 35 & 104637.3 & 15.0 & Sm & 25 & 1384 & 1 & $1.\arcmin9$ & $6.\arcmin7$ & 6.1 \nl
 &  & 124104 & 1.04 & FS & 1.6 & 74 & 2 & $1.\arcmin6$ & $\sim 7\arcmin$ & 1.7 \nl
 & & & & & & & & & & \nl
14 & FS 36 & 104638.8 & 17.0 & ImIV & 12 & 1345 & 1 & $1.\arcmin9$ & $4.\arcmin1$ & 1.6 \nl
 &  & 123824 & 0.89 & FS & 1.4 & 25 & 1 & $1.\arcmin2$ & $\lesssim4\arcmin$ & 1.7 \nl
 & & & & & & & & & & \nl
15 & BTS 27 & 105801.1 & 15.5 & ImIII & 24 & 2392 & 1 & $1.\arcmin9$ & $< 1.\arcmin9$ & 4.1 \nl
 &  & 292242 & 0.88 & BTS & 1.1 & 70 & 2 & $1.\arcmin1$ & $\lesssim4\arcmin$ & 1.6 \nl
 & & & & & & & & & & \nl
16 & BTS 29 & 105855.7 & 16.5 & SmIV pec & 16 & 1680 & 1 & $1.\arcmin9$ & $< 1.\arcmin9$ & 5.6 \nl
 &  & 305236 & 1.27 & BTS & 1.2 & 71 & 2 & $2.\arcmin8$ & $\lesssim4\arcmin$ & 1.6 \nl
 & & & & & & & & & & \nl
17 & A1122+1709 & 112233.7 & 15.0 & SmIII & 17 & 1023 & 2 & $1.\arcmin9$ & $< 1.\arcmin9$ & 0.7 \nl
 &  & 170936 & 0.64 & AS & 1.1 & 32 & 4 & $0.\arcmin6$ & $< 4\arcmin$ & 1.8 \nl
 & & & & & & & & & & \nl
18 & DDO 97 & 114619.5 & 14.76 & ImV & 11 & 523 & 1 & $1.\arcmin9$ & $4.\arcmin2$ & 8.3 \nl
 & UGC 6782 & 240700 & 1.31 & RC2 & 1.3 & 88 & 1 & $3.\arcmin1$ & $\sim5\arcmin$ & 1.5 \nl
 & & & & & & & & & & \nl
19 & DDO 101 & 115306.0 & 14.5 & Sd & 15 & 590 & 1 & $1.\arcmin9$ & $2.\arcmin0$ & 2.1 \nl
 & UGC 6900 & 314742 & 1.32 & BTS & 1.3 & 100 & 2 & $3.\arcmin1$ &  & \nl
 & & & & & & & & & & \nl
20 & A1156+1409 & 115659.0 & 14.43 & SmIII & 26 & 1448 & 1 & $1.\arcmin9$ & $2.\arcmin1$ & 5.0 \nl
 &  & 140948 & 0.77 & AS & 1.0 & 56 & 2 & $0.\arcmin9$ & $< 4\arcmin$ & 3.6 \nl
 & & & & & & & & & & \nl
21 & BTS 89 & 115914.2 & 16.5 & Im & 12 & 842 & 1 & $1.\arcmin9$ & $< 1.\arcmin9$ & 0.5 \nl
 &  & 283812 & 0.91 & BTS & 1.1 & 29 & 1 & $1.\arcmin2$ & $\lesssim4\arcmin$ & 1.6 \tablebreak
 & & & & & & & & & & \nl
22 & VCC 10 & 120651.6 & 14.75 & BCD & 19 & 1972 & 3 & $1.\arcmin9$ & $< 1.\arcmin9$ & 2.4 \nl
 & IC 3017 & 135106 & 0.89 & BST & 1.1 & 170 & 5 & $1.\arcmin2$ & $< 4\arcmin$ & 2.0 \nl
 & & & & & & & & & & \nl
23 & IC 3023 & 120729.0 & 15.2 & ImIV & 21 & 820 & 2 & $2.\arcmin5$ & $< 1.\arcmin9$ & 3.1 \nl
 & UGC 7150 & 143824 & 1.11 & BST & 1.1 & 48 & 4 & $1.\arcmin9$ & $< 4\arcmin$ & 1.3 \nl
 & & & & & & & & & & \nl
24 & VCC 26 & 120807.1 & 17.5 & Im? & 13 & 2467 & 1 & $1.\arcmin9$ & $2.\arcmin8$ & 1.5 \nl
 &  & 145530 & 0.74 & BST & 1.4 & 38 & 1 & $0.\arcmin8$ & $\lesssim4\arcmin$ & 2.2 \nl
 & & & & & & & & & & \nl
25 & NGC 4186 & 121053.0 & 15.6 & d:Sc & 14 & 2076 & 1 & $1.\arcmin9$ & $2.\arcmin2$ & 5.7 \nl
 & UGC 7223 & 150300 & 1.19 & BST & 1.1 & 82 & 1 & $2.\arcmin3$ & $\sim 4\arcmin$ & 1.5 \nl
 & & & & & & & & & & \nl
26 & IC 3049 & 121100.0 & 15.13 & ImIII-IV & 16 & 2443 & 1 & $2.\arcmin5$ & $2.\arcmin5$ & 2.4 \nl
 & UGC 7227 & 144530 & 1.00 & BST & 1.2 & 106 & 2 & $1.\arcmin5$ &  & \nl
 & & & & & & & & & & \nl
27 & VCC 114 & 121201.2 & 16.0 & Im? & 14 & 2068 & 2 & $1.\arcmin9$ & $2.\arcmin8$ & 2.1 \nl
 &  & 55718 & 0.89 & BST & 1.2 & 31 & 3 & $1.\arcmin2$ & $\lesssim4\arcmin$ & 3.8 \nl
 & & & & & & & & & & \nl
28 & BTS 105 & 121233.0 & 13.9 & BCD & 24 & 956 & 5 & $1.\arcmin9$ & $3.\arcmin9$ & 18.2 \nl
 & UGC 7257 & 361400 & 1.18 & BTS & 1.2 & 89 & 9 & $2.\arcmin3$ & $\sim5\arcmin$ & 2.0 \nl
 & & & & & & & & & & \nl
29 & VCC 159 & 121308.0 & 15.08 & ImIII & 26 & 2581 & 2 & $1.\arcmin9$ & $3.\arcmin0$ & 2.7 \nl
 &  & 83347 & 0.93 & BST & 1.1 & 47 & 4 & $1.\arcmin3$ & $\lesssim4\arcmin$ & 3.6 \nl
 & & & & & & & & & & \nl
30 & VCC 169 & 121323.4 & 16.5 & ImV & 11 & 2220 & 1 & $1.\arcmin9$ & $2.\arcmin4$ & 6.0 \nl
 & $9\arcdeg5$ & 95536 & 1.00 & BST & 1.2 & 22 & 1 & $1.\arcmin5$ & $\lesssim4\arcmin$ & 3.1 \nl
 & & & & & & & & & & \nl
31 & UGC 7307 & 121432.0 & 15.5 & ImIV-V: & 17 & 1183 & 1 & $1.\arcmin9$ & $3.\arcmin5$ & 5.2 \nl
 & VCC 217 & 101700 & 1.25 & BST & 1.1 & 60 & 2 & $2.\arcmin7$ & $\lesssim4\arcmin$ & 1.2 \nl
 & & & & & & & & & & \nl
32 & IC 3142a & 121633.0 & 15.1 & Im: & 37 & $-209$ & 2 & $1.\arcmin9$ & $3.\arcmin0$ & 3.4 \nl
 & UGC 7355 & 141518 & 1.00 & BST & 1.1 & 67 & 4 & $1.\arcmin5$ & $< 4\arcmin$ & 3.5 \nl
 & & & & & & & & & & \nl
33 & UGC 7354 & 121636.6 & 14.78 & BCD & 20 & 1525 & 2 & $1.\arcmin9$ & $< 1.\arcmin9$ & 2.8 \nl
 & VCC 324 & 40800 & 0.81 & BST & 1.1 & 47 & 3 & $1.\arcmin0$ & $< 4\arcmin$ & 3.8 \nl
 & & & & & & & & & & \nl
34 & VCC 340 & 121648.6 & 14.43 & BCD or merger & 33 & 1515 & 2 & $1.\arcmin9$ & $2.\arcmin2$ & 3.6 \nl
 &  & 61130 & 0.93 & BST & 1.1 & 49 & 5 & $1.\arcmin3$ & $< 4\arcmin$ & 3.2 \nl
 & & & & & & & & & & \nl
35 & VCC 350 & 121654.6 & 17.2 & ImIV-V: & 9 & 304 & 1 & $1.\arcmin9$ & $< 1.\arcmin9$ & 1.5 \nl
 & $13\arcdeg30$ & 133512 & 0.90 & BST & 1.0 & 28 & 2 & $1.\arcmin2$ & $< 4\arcmin$ & 2.8 \nl
 & & & & & & & & & & \nl
36 & VCC 381 & 121720.0 & 16.5 & ImV & 15 & 481 & 1 & $1.\arcmin9$ & $3.\arcmin8$ & 3.2 \nl
 & $6\arcdeg18$ & 65635 & 1.00 & BST & 1.3 & 24 & 1 & $1.\arcmin5$ & $\lesssim4\arcmin$ & 1.8 \nl
 & & & & & & & & & & \nl
37 & VCC 459 & 121839.0 & 14.95 & BCD & 28 & 2099 & 5 & $2.\arcmin5$ & $4.\arcmin2$ & 2.4 \nl
 &  & 175454 & 0.80 & BST & 1.2 & 110 & 10 & $0.\arcmin9$ &  & \nl
 & & & & & & & & & & \nl
38 & VCC 566 & 122005.0 & 15.8 & SBmIII: & 17 & 1408 & 2 & $2.\arcmin5$ & $4.\arcmin1$ & 3.9 \nl
 &  & 83424 & 0.93 & BST & 1.2 & 85 & 4 & $1.\arcmin3$ &  & \nl
 & & & & & & & & & & \nl
39 & VCC 826 & 122308.0 & 15.0 & ImIV & 17 & 1506 & 2 & $2.\arcmin5$ & $4.\arcmin8$ & 7.0 \nl
 & UGC 7512 & 22605 & 1.12 & BST & 1.4 & 65 & 3 & $2.\arcmin0$ &  & \nl
 & & & & & & & & & & \nl
40 & VCC 848 & 122319.7 & 14.72 & ImIII pec/BCD & 21 & 1542 & 5 & $1.\arcmin9$ & $3.\arcmin7$ & 8.2 \nl
 &  & 60506 & 1.11 & BST & 1.2 & 138 & 10 & $1.\arcmin9$ & $\sim 5\arcmin$ & 0.9 \nl
 & & & & & & & & & & \nl
41 & IC 3356 & 122421.0 & 14.49 & SmIV & 12 & 1105 & 1 & $1.\arcmin9$ & $3.\arcmin7$ & 19.9 \nl
 & UGC 7547 & 115018 & 1.22 & BST & 1.2 & 75 & 3 & $2.\arcmin5$ & $\sim 7\arcmin$ & 1.3 \tablebreak
 & & & & & & & & & & \nl
42 & UGC 7642 & 122740.0 & 14.63 & SdmIII-IV & 17 & 1638 & 1 & $2.\arcmin5$ & $4.\arcmin4$ & 5.1 \nl
 & VCC 1266 & 25406 & 1.07 & BST & 1.3 & 44 & 2 & $1.\arcmin8$ &  & \nl
 & & & & & & & & & & \nl
43 & IC 3446 & 122851.6 & 14.9 & SmIII/BCD & 14 & 1230 & 2 & $1.\arcmin9$ & $2.\arcmin4$ & 4.4 \nl
 & VCC 1356 & 114600 & 0.93 & BST & 1.2 & 153 & 4 & $1.\arcmin3$ & $\lesssim4\arcmin$ & 3.6 \nl
 & & & & & & & & & & \nl
44 & DDO 131 & 122928.1 & 15.0 & ImIII-IV & 10 & 645 & 1 & $1.\arcmin9$ & $4.\arcmin0$ & 13.9 \nl
 & UGC 7673 & 295854 & 1.20 & BTS & 1.2 & 70 & 1 & $2.\arcmin4$ & $\sim5\arcmin$ & 1.9 \nl
 & & & & & & & & & & \nl
45 & VCC 1468 & 123024.6 & 15.0 & ImIV & 10 & 1230 & 2 & $1.\arcmin9$ & $2.\arcmin4$ & 3.6 \nl
 &  & 45112 & 1.06 & BST & 1.2 & 85 & 4 & $1.\arcmin7$ &  & \nl
 & & & & & & & & & & \nl
46 & IC 3522 & 123215.0 & 15.2 & ImIII-IV pec & 19 & 664 & 1 & $1.\arcmin9$ & $\lesssim1.\arcmin9$ & 10.7 \nl
 & UGC 7737 & 152948 & 1.25 & BST & 1.2 & 105 & 2 & $2.\arcmin7$ & $\lesssim4\arcmin$ & 1.5 \nl
 & & & & & & & & & & \nl
47 & VCC 1644 & 123321.0 & 17.5 & SmIV & 24 & 764 & 1 & $1.\arcmin9$ & $2.\arcmin0$ & 2.3 \nl
 &  & 140800 & 1.05 & BST & 1.1 & 72 & 2 & $1.\arcmin7$ &  & \nl
 & & & & & & & & & & \nl
48 & VCC 1654 & 123338.0 & 15.04 & ImIII & 29 & 2056 & 2 & $1.\arcmin9$ & $3.\arcmin2$ & 3.1 \nl
 & IC 3562 & 101200 & 1.00 & BST & 1.2 & 67 & 4 & $1.\arcmin5$ & $\sim 4\arcmin$ & 2.2 \nl
 & & & & & & & & & & \nl
49 & IC 3583 & 123413.0 & 13.91 & SmIII & 19 & 1121 & 1 & $1.\arcmin9$ & $< 1.\arcmin9$ & 6.5 \nl
 & UGC 7784 & 133200 & 1.32 & BST & 1.1 & 107 & 2 & $3.\arcmin1$ & $\sim 4\arcmin$ & 1.6 \nl
 & & & & & & & & & & \nl
50 & IC 3589 & 123430.0 & 14.11 & SBmIII & 26 & 1636 & 6 & $2.\arcmin5$ & $3.\arcmin0$ & 10.1 \nl
 & UGC 7790 & 71217 & 1.11 & BST & 1.2 & 109 & 12 & $1.\arcmin9$ &  & \nl
 & & & & & & & & & & \nl
51 & VCC 1725 & 123509.0 & 14.51 & SmIII/BCD & 37 & 1073 & 4 & $2.\arcmin5$ & $2.\arcmin4$ & 2.6 \nl
 &  & 85000 & 1.11 & BST & 1.1 & 78 & 8 & $1.\arcmin9$ &  & \nl
 & & & & & & & & & & \nl
52 & VCC 1726 & 123513.0 & 14.54 & SdmIV & 31 & 63 & 2 & $2.\arcmin5$ & $2.\arcmin4$ & 7.1 \nl
 & UGC 7795 & 72247 & 1.14 & BST & 1.2 & 71 & 4 & $2.\arcmin1$ &  & \nl
 & & & & & & & & & & \nl
53 & BTS 145 & 123613.7 & 16.0 & ImIV-V & 12 & 309 & 1 & $1.\arcmin9$ & $4.\arcmin0$ & 20.5 \nl
 & CVn dwA & 330230 & 1.31 & BTS & 1.2 & 30 & 1 & $3.\arcmin1$ & $\sim6\arcmin$ & 1.8 \nl
 & & & & & & & & & & \nl
54 & IC 3617 & 123653.0 & 14.67 & SBmIII/BCD & 44 & 2084 & 2 & $2.\arcmin5$ & $4.\arcmin0$ & 8.6 \nl
 & UGC 7822 & 81412 & 1.26 & BST & 1.2 & 91 & 4 & $2.\arcmin7$ &  & \nl
 & & & & & & & & & & \nl
55 & VCC 1816 & 123728.0 & 16.2 & ImIV-V & 12 & 1008 & 1 & $1.\arcmin9$ & $< 1.\arcmin9$ & 3.7 \nl
 & $14\arcdeg63$ & 140324 & 1.11 & BST & 1.1 & 31 & 1 & $1.\arcmin9$ & $\sim 5\arcmin$ & 1.5 \nl
 & & & & & & & & & & \nl
56 & VCC 1952 & 124034.8 & 16.0 & ImIV & 20 & 1313 & 1 & $1.\arcmin9$ & $< 1.\arcmin9$ & 2.8 \nl
 & $7\arcdeg50$ & 75524 & 0.93 & BST & 1.1 & 53 & 2 & $1.\arcmin3$ & $\lesssim4\arcmin$ & 2.9 \nl
 & & & & & & & & & & \nl
57 & VCC 1992 & 124139.0 & 15.5 & ImIV & 18 & 1008 & 2 & $1.\arcmin9$ & $< 1.\arcmin9$ & 6.2 \nl
 & UGC 7906 & 122324 & 1.11 & BST & 1.1 & 107 & 4 & $1.\arcmin9$ & $\lesssim4\arcmin$ & 2.0 \nl
 & & & & & & & & & & \nl
58 & VCC 2062 & 124528.8 & 19.0: & dE: & 24 & 1146 & 4 & $1.\arcmin9$ & $3.\arcmin7$ & 7.2 \nl
 & $11\arcdeg70$ & 111454 & 0.83 & BST & 1.4 & 46 & 8 & $1.\arcmin0$ & $\sim 5\arcmin$ & 1.4 \nl
 & & & & & & & & & & \nl
59 & NGC 4694 & 124543.8 & 12.19 & Amorphous & 34 & 1184 & 4 & $1.\arcmin9$ & $2.\arcmin1$ & 3.7 \nl
 & UGC 7969 & 111524 & 1.56 & BST & 1.2 & 81 & 8 & $5.\arcmin4$ & $\lesssim4\arcmin$ & 1.2 \nl
 & & & & & & & & & & \nl
60 & DDO 154 & 125139.5 & 14.39 & SmIV & 14 & 375 & 2 & $3.\arcmin8$ & $9.\arcmin2$ & 129.5 \nl
 & UGC 8024 & 272518 & 1.26 & CB & 1.8 & 88 & 5 & $2.\arcmin7$ & $25.\arcmin6$ & 0.6 \nl
 & & & & & & & & & & \nl
61 & UGC 8055 & 125331.1 & 17. & ImV & 10 & 617 & 1 & $1.\arcmin9$ & $2.\arcmin7$ & 8.0 \nl
 &  & 40454 & 1.06 & UGC & 1.1 & 89 & 2 & $1.\arcmin7$ & $\sim4\arcmin$ & 1.6 \tablebreak
 & & & & & & & & & & \nl
62 & DDO 155 & 125609.2 & 14.82 & ImV & 15 & 215 & 1 & $1.\arcmin9$ & $3.\arcmin2$ & 8.6 \nl
 & UGC 8091 & 142913 & 1.08 & CBF & 1.2 & 30 & 1 & $1.\arcmin8$ & $\lesssim4\arcmin$ & 1.4 \nl
 & & & & & & & & & & \nl
63 & $13\arcdeg118$ & 125623.1 & 14.78 & dIrr(I/II) & 25 & 1901 & 1 & $1.\arcmin9$ & $3.\arcmin5$ & 6.5 \nl
 &  & 132520 & 1.08 & GR & 1.2 & 57 & 2 & $1.\arcmin8$ & $\lesssim4\arcmin$ & 1.9 \nl
 & & & & & & & & & & \nl
64 & $13\arcdeg118$a & 1256.4 &  &  &  & 1945 & 2 & $1.\arcmin9$ & $< 1.\arcmin9$ & 0.9 \nl
 &  & 1323 & 0.5 & POSS &  & 43 & 4 & $0.\arcmin5$ & $< 4\arcmin$ & 1.9 \nl
 & & & & & & & & & & \nl
65 & UGC 8114 & 125755.3 & 15.4 & pec & 26 & 1991 & 1 & $1.\arcmin9$ & $2.\arcmin5$ & 8.7 \nl
 & $13\arcdeg120$ & 135613 & 1.22 & UGC & 1.2 & 132 & 2 & $2.\arcmin5$ & $\lesssim4\arcmin$ & 1.5 \nl
 & & & & & & & & & & \nl
66 & BTS 163 & 130209.0 & 16.5 & Im? & 17 & 1932 & 1 & $1.\arcmin9$ & $2.\arcmin8$ & 4.0 \nl
 &  & 270212 & 0.91 & BTS & 1.2 & 95 & 2 & $1.\arcmin2$ & $\lesssim4\arcmin$ & 1.5 \nl
 & & & & & & & & & & \nl
67 & DDO 187 & 141337.7 & 14.35 & ImIV-V & 12 & 152 & 1 & $1.\arcmin9$ & $3.\arcmin6$ & 14.2 \nl
 & UGC 9128 & 231706 & 1.25 & T & 1.1 & 36 & 2 & $2.\arcmin7$ & $\sim5\arcmin$ & 1.2 \nl
 & & & & & & & & & & \nl
68 & BTS 171 & 143107.7 & 15.5 & BCD & 7 & 835 & 1 & $1.\arcmin9$ & $< 1.\arcmin9$ & 2.4 \nl
 &  & 271242 & 1.01 & BTS & 1.1 & 78 & 1 & $1.\arcmin5$ &  & \nl
 & & & & & & & & & & \nl
69 & BTS 177 & 153509.0 & 15.5 & SBd & 13 & 1868 & 1 & $1.\arcmin9$ & $2.\arcmin5$ & 7.6 \nl
 & UGC 9938 & 301436 & 1.20 & BTS & 1.1 & 99 & 1 & $2.\arcmin4$ & $\sim4\arcmin$ & 2.0 \nl
 & & & & & & & & & & \nl
70 & DDO 216 & 232602.7 & 12.63 & ImV & 88 & $-189$ & 1 & $1.\arcmin9$ & $6.\arcmin9$ & 22.8 \nl
 & UGC 12613 & 142813 & 1.66 & T & 1.2 & 32 & 1 & $6.\arcmin9$ & $\sim8\arcmin$ & 1.0 \nl
\enddata
\end{deluxetable}